\documentclass[%
 superscriptaddress,
 twocolumn,
nofootinbib,
 amsmath,amssymb,
 aps,
 prc,
]{revtex4-2}

\usepackage[colorlinks=true,allcolors=blue]{hyperref} 
\usepackage{graphicx}
\usepackage{dcolumn}
\usepackage{bm}
\usepackage{physics} 
\renewcommand{\vec}[1]{\boldsymbol{#1}} 
\newcolumntype{d}[1]{D{.}{.}{#1}} 
\usepackage{xcolor}
\usepackage{soul}
\usepackage{amsthm}

\usepackage{ulem}

\begin{document}


\title{The axially-deformed relativistic quasiparticle random phase approximation based on point-coupling interactions}

\author{A. Ravli\'c}
\email[]{ravlic@frib.msu.edu}
\affiliation{Facility for Rare Isotope Beams, Michigan State University, East Lansing, Michigan 48824, USA}

\author{T. Nik\v{s}i\'c}
\email[]{tniksic@phy.hr}
\affiliation{Department of Physics, Faculty of Science, University of Zagreb, Bijeni\v{c}ka c. 32, 10000 Zagreb, Croatia}

\author{Y. F. Niu}
\email[]{niuyf@lzu.edu.cn}
\affiliation{School of Nuclear Science and Technology, Lanzhou University, Lanzhou 730000, China}

\author{P. Ring}
\affiliation{Physik Department, Technische Universit\"at M\"unchen, D-85747 Garching, Germany}

\author{N. Paar}
\email[]{npaar@phy.hr}
\affiliation{Department of Physics, Faculty of Science, University of Zagreb, Bijeni\v{c}ka c. 32, 10000 Zagreb, Croatia}

\date{\today}

\begin{abstract}
Collective nuclear excitations, like giant resonances, are
sensitive to nuclear deformation, as evidenced
by alterations in their excitation energies and transition strength distributions.
A common theoretical framework to study these collective modes, the random-phase approximation (RPA), has to deal with large dimensions spanned by all possible particle-hole configurations satisfying certain symmetries. 
It is the aim of this work to establish a new theoretical framework to study the impact of deformation on spin-isospin excitations, that is able to provide fast and reliable solutions of the RPA equations. 
The nuclear ground state is determined with the axially-deformed relativistic Hartree-Bogoliubov (RHB) model based on relativistic point-coupling energy density functionals (EDFs). To study the excitations in the charge-exchange channel, an axially-deformed proton-neutron relativistic quasiparticle RPA (pnRQRPA) is developed in the linear response approach.
After benchmarking the axially-deformed pnRQRPA in the spherical limit, a study of spin-isospin excitations including Fermi, Gamow-Teller (GT) and Spin-Dipole (SD) is performed for selected $pf$-shell nuclei. For GT transitions, it is demonstrated that deformation leads to a considerable fragmentation of the strength function. A mechanism inducing the fragmentation is studied by decomposing the total strength to different projections of total angular momentum $K$ and constraining the nuclear shape to either spherical, prolate or oblate. A similar fragmentation is also observed for SD transitions, although somewhat moderated by the complex structure of these transitions, while, as expected, the Fermi strength is almost shape-independent. The axially-deformed pnRQRPA introduced in this work open perspectives for the future studies of deformation effects on astrophysically relevant weak interaction processes, in particular beta decay and electron capture.
\end{abstract}

\maketitle

\section{Introduction}

The spin-isospin excitation is one of the fundamental collective modes of a nucleus. Similar to the magnetic excitations ($M\lambda$), they induce spin-transitions between different nuclear states, however, they also have the isospin component, which allows proton-neutron mixing. According to the total angular momentum $J$ and parity $\pi$, spin-isospin excitations are categorized as Fermi ($0^+$), Gamow-Teller ($1^+$), spin-dipoles ($0^-, 1^-, 2^-$) and other higher order multipoles. They play a major role in determining the abundance patterns of $r$-process nucleosynthesis through $\beta$-decay rates \cite{KAJINO2019109,MUMPOWER201686}, as well as the dynamics of core-collapse supernovae through electron captures \cite{Langanke_2021}. Spin-isospin transitions also have significant implications in the field of fundamental symmetries by playing a major role in determining the nuclear matrix elements of the neutrinoless double $\beta$-decay, which, if measured, would prove the existence of Majorana neutrinos and physics beyond the standard model \cite{Engel_2017,YAO2022103965}. Furthermore, the difference between isobaric analog and Gamow-Teller (GT) resonance energies can be used to extract the neutron skin thickness, an important quantity closely related to the symmetry energy $J$ of infinite nuclear matter \cite{PhysRevLett.91.262502,ROCAMAZA201896}. Additionally, the neutron skin-thickness can also be extracted from the sum rules of spin-dipole (SD) excitations \cite{PhysRevLett.82.3216}. Therefore, it is of great theoretical interest to further improve our understanding of spin-isospin excitations.

Currently, there is a plethora of theoretical models with the ability to investigate spin-isospin excitations, which can be grouped into three main categories: (i) ab initio approaches, (ii) the nuclear shell-model, and (iii) energy density functional theory (EDF).
 Although there is significant progress in recent years in 
ab initio nuclear theory \cite{PhysRevLett.124.232501,PhysRevLett.127.242502,PhysRevLett.126.042502,PhysRevC.106.014315,Gysbers2019}, based on nuclear interactions derived from chiral effective field theory, the study of spin-isospin modes is restricted to considering specific nuclei only. The shell model can handle the description of spin-isospin excitations and resonances up to the tin region of the nuclide chart, after which the dimension of the Hamiltonian becomes too large even for modern-day computers \cite{RevModPhys.77.427}. On the other hand, there is no other model with such excellent scaling property as the EDF theory \cite{RevModPhys.75.121,NIKSIC2011519}, which, in principle, allows for the calculation of spin-isospin transitions throughout the nuclide chart, from the proton to the neutron drip line. Since ultimately, especially for astrophysical applications, it is necessary to describe spin-isospin excitations globally, the EDF theory is currently the only theoretical approach that allows for such endeavors.

In EDF theory the energy is expressed as a functional of nuclear density $\rho$, and its minimization leads to the nuclear ground state. In principle, this is an exact theory if the underlying functional corresponds to the exact nucleon-nucleon interaction. Since such approaches are still unfeasible, nuclear EDF theory often resorts to phenomenological functionals. These phenomenological functionals are separated into two categories: (i) non-relativistic and (ii) relativistic. In this work, we focus on relativistic EDFs, where nucleons are point-like Dirac particles, and their effective interaction is described with four-fermion contact interaction terms, including isoscalar-scalar, isoscalar-vector, and isovector-vector channels \cite{PhysRevC.78.034318,NIKSIC20141808}. The corresponding Lagrangian density includes free nucleon terms, point coupling interaction terms, coupling of protons to the electromagnetic field, and the derivative term accounting for the leading effects of finite-range interactions \cite{PhysRevC.78.034318}. For a quantitative description of nuclear density distributions and radii, the derivative terms are necessary \cite{PhysRevC.78.034318}. The model includes the density dependence explicitly through couplings in the 
interactions terms \cite{NIKSIC20141808}.
The relativistic EDF allows the description of the coordinate and spin degrees of freedom on the same footing, enabling the natural inclusion of spin-orbit terms without any extra parameters. To consider superfluid nuclei one has to transform from the basis of single-particle Dirac states to quasi-particle (q.p.) states using the Bogoliubov transformation, leading to the relativistic Hartree Bogoliubov (RHB) equation which determines the ground-state of a superfluid nucleus \cite{VRETENAR2005101}.

To build the nuclear excitations on top of the RHB ground state, we can consider an excitation operator comprised of a superposition of 2 q.p. states, which leads to the theory known as the relativistic quasiparticle random-phase approximation (RQRPA) \cite{PhysRevC.67.034312}. If the 2 q.p. states consist of a proton-neutron pair, then we can construct the proton-neutron RQRPA (pnRQRPA), which allows for a description of spin-isospin excitations \cite{PhysRevC.69.054303}. The corresponding non-relativistic version is termed pnQRPA \cite{PhysRevC.60.014302}.

In our previous work \cite{PhysRevC.104.064302}, we demonstrated the basics of the pnRQRPA equations in the response function formalism assuming spherical symmetry. In this case it is advantageous to use the fact that equations decouple based on angular momentum and parity $J^\pi$ blocks once the proper angular momentum coupling is performed. The angular momentum coupling leads to a reasonable dimension of the model space that requires only moderate computational resources. We have applied spherical pnRQRPA to study spin-isospin excitations in tin isotopes. Since these nuclei have closed proton shells at $Z = 50$ and are located near the $N = 82$ neutron magic number, they are spherical or near-spherical in their ground state. However, for both proton and neutron open shells, most nuclei discovered so far are axially deformed, requiring an extension of the model. The calculations in the axial geometry are significantly more complicated than the spherical calculations. First of all, compared to the spherical geometry, the dimension of the model space is much larger since no angular momentum coupling to $J^\pi$ blocks can be performed. However, in axial deformation, the angular momentum projection on the $z-$axis, $K$, is still a good quantum number, and if we also consider reflection-symmetric shapes, the pnRQRPA equations can be decoupled into $K^\pi$ blocks. Instead of diagonalizing the pnRQRPA matrix in the space of all 2 q.p. pairs, which is around $10^5$ for a reasonable basis size, we employ the linear response formalism built for point-coupling EDFs with separable pairing interactions \cite{PhysRevC.78.034318,PhysRevC.99.034318}. This allows us to work in the space determined by the number of interaction channels and the coordinate mesh, which is around $10^3$, significantly lower than the 2 q.p. dimension.

Due to high computational costs, most of previous calculations of charge-exchange transitions for deformed nuclei assume schematic models, usually with simple separable interactions. Such models have been applied to study e.g., the rotational excitations \cite{RING1984261}, spin-isospin excitations, and subsequent calculations of $\beta$-decay half-lives \cite{PhysRevC.91.044304,SARRIGUREN199855,SARRIGUREN2001631}, including the double $\beta$-decay in both 2$\nu$ \cite{PhysRevC.70.064309,SIMKOVIC2004321} and neutrinoless channels \cite{SIMKOVIC1996257}. Although based on relatively simple interactions, such calculations led to important discoveries concerning the impact of deformation on observables of interest. As computing power and numerical techniques were improved, more sophisticated implementations of the pn(R)QRPA based on the EDF theory have been developed. Most of these are constrained to spherical nuclei \cite{PhysRevLett.101.122502,PhysRevC.95.044301,PhysRevC.103.064307,PhysRevC.69.054303,PhysRevC.60.014302}. Concerning the method for solving pn(R)QRPA equations, on the one hand, there are approaches based on the matrix equations, which require diagonalizing large matrices \cite{PhysRevC.77.034317,Yoshida_ptep,PhysRevC.78.064316,PhysRevC.83.021304,PhysRevC.94.014304}. On the other hand, implementations based on the finite-amplitude method (FAM), where one avoids the diagonalization and solves the equations of motion for each excitation energy iteratively, were presented in Refs. \cite{PhysRevC.76.024318,PhysRevC.84.014314}. The FAM method has also been applied to spin-isospin excitations but only with non-relativistic EDFs in Refs. \cite{PhysRevC.90.024308,PhysRevC.93.014304,PhysRevC.102.034326}. Concerning the relativistic EDFs, approaches based on the FAM were developed in Refs. \cite{BJELCIC2020107184,PhysRevC.88.044327} and used to obtain the electric response of axially-deformed nuclei \cite{PhysRevC.103.024303}. To date, there are no calculations with relativistic EDFs applied to spin-isospin excitations in deformed nuclei. Therefore, in this work we develop the pnRQRPA in axial geometry based on the relativistic EDFs. Furthermore, we aim to develop a pnRQRPA solver with comparable speed as the FAM, and propose an alternative to the variational approach.

This work is organized as follows. First, we introduce the axially-deformed pnRQRPA formalism and present the calculation techniques for external field and residual interaction matrix elements in Sec. \ref{sec:theory}. This is supplemented with appendices \ref{sec:appa} and \ref{sec:appb}, where additional details about numerical implementations are given. Secondly, we perform numerical tests of our axially-deformed pnRQRPA by comparing it to the spherical pnRQRPA from Ref. \cite{PhysRevC.104.064302} in Sec. \ref{sec:numerical_tests}. After properly testing the model calculations and determining the optimal basis size for the nuclei of interest, we present the calculations of the Gamow-Teller, Fermi, and spin-dipole transitions of particular even-even $pf$-shell nuclei in Sec. \ref{sec:AXDEF_spin_isospin}. 

\section{Theoretical formalism}\label{sec:theory}

\subsection{Brief introduction to the linear response pnRQRPA}
The relativistic Hartree-Bogoliubov (RHB) model \cite{Kucharek1991,VRETENAR2005101} provides a unified description of nuclear particle-hole ($ph$) and particle-particle ($pp$) correlations on a mean-field level by combining two average potentials: the consistent nuclear mean-field $h$ that includes all the long range $ph$ correlations, and a pairing field 
$\Delta$ which sums up the $pp$ correlations. In the RHB framework the nuclear single-reference state is described by a generalized Slater determinant $|\Phi \rangle$ that represents a vacuum with respect to independent quasiparticles. The quasiparticle operators are defined by the unitary Bogoliubov transformation, and the corresponding Hartree-Bogoliubov wave functions $U$ and $V$ are determined by the solution of the RHB equation
\begin{equation}
\left( \begin{array}{cc} h_D-m-\lambda & \Delta \\ -\Delta^* & 
-h_D^*+m+\lambda \end{array} \right) 
\left(\begin{array}{c} U_\mu \\ V_\mu \end{array} \right)
=E_\mu \left(\begin{array}{c} U_\mu \\ V_\mu \end{array} \right).
\end{equation}
The original single-particle basis $c_k^\dag, c_k$ (e.g. a harmonic oscillator basis) is transformed to the quasparticle (q.p.) basis 
$ \beta_\mu, \beta_\mu^\dag $ by the Bogoliubov transformation \cite{ring2004nuclear}
\begin{align}
    \beta_\mu^\dag = \sum \limits_{l = 1}^M U_{l \mu} c_l^\dag + V_{l \mu} c_l, \\
    \beta_\mu = \sum \limits_{l = 1}^M V_{l \mu}^* c_l^\dag + U_{l\mu}^* c_l,
\end{align}
where $M$ denotes the dimension of the single-particle basis. 
We introduce the $2M$ dimensional set of extended q.p. states $a_\mu$, which conveniently combines the q.p. operators $\beta_\mu, \beta_\mu^\dag$ as \cite{RING1984261}

\begin{equation}\label{eq:doubled_basis}
    \left. \begin{array}{c}
         a_\mu = \beta_\mu  \\
         a_{\tilde{\mu}} = \beta_\mu^\dag 
    \end{array} \right\}, \quad \mu=1,\dots,M;\;\; \tilde{\mu}=-\mu ,
\end{equation}
which obey the anticommutation relation $\{a_\mu, a_{\mu^\prime} \} = \delta_{\mu \tilde{\mu}^\prime}$. The main idea behind the transformation (\ref{eq:doubled_basis}) is to represent both the creation and the annihilation operators using one set of operators $a_\mu$, with  $a_\mu = \beta_\mu$ for $\mu > 0$ and $a_\mu = \beta_{-\mu}^\dag$ for $\mu < 0$. In this basis, one-body q.p. operator is represented as \cite{RING1984261}
\begin{equation}
    \hat{F} = \hat{F}^0 + \frac{1}{2} \sum \limits_{\mu \mu^\prime} \mathcal{F}_{\mu \mu^\prime} a_\mu^\dag a_{\mu^\prime}.
\end{equation}
The matrix elements $\mathcal{F}_{\mu \mu^\prime}$ with signs of the $\mu$ and $\mu^\prime$ indices explicitly specified read
\begin{equation}\label{eq:signature_2qp}
    \mathcal{F}_{\mu \mu^\prime} = \begin{pmatrix}
        \mathcal{F}_{\mu > 0 \mu^\prime > 0} & \mathcal{F}_{\mu > 0 \mu^\prime < 0} \\
        \mathcal{F}_{\mu < 0 \mu^\prime > 0} & \mathcal{F}_{\mu < 0 \mu^\prime < 0}
    \end{pmatrix} .
\end{equation}
Furthermore, the following relations can easily be verified
\begin{align}
    \begin{split}
        \mathcal{F}_{\mu > 0 \mu^\prime > 0} &= F^{11}_{\mu \mu^\prime} , \\
       \mathcal{F}_{\mu > 0 \mu^\prime <0} &=  F^{20}_{\mu -\mu^\prime} , \\
    \mathcal{F}_{\mu < 0 \mu^\prime > 0} &= - F^{02}_{-\mu \mu^\prime} , \\
\mathcal{F}_{\mu < 0 \mu^\prime < 0} &=- F^{\bar{11}}_{-\mu -\mu^\prime} , \\
    \end{split}
\end{align}
where $F^{11}, F^{20}, F^{02}$, and $F^{\bar{11}}$ are the q.p. components of the one-body operator defined in Ref. \cite{ring2004nuclear}.

The $2M \times 2M$ generalized density matrix $\mathcal{R}$ acquires the following form
\begin{equation}
    \mathcal{R} = \begin{pmatrix}
        \langle \Phi | a_{\tilde{\mu}^\prime} a_\mu | \Phi \rangle &  \langle \Phi | a_{\mu^\prime} a_\mu | \Phi \rangle \\
        \langle \Phi | a_{\tilde{\mu}^\prime} a_{\tilde{\mu}} | \Phi \rangle &  \langle \Phi | a_{\mu^\prime} a_{\tilde{\mu}} | \Phi \rangle \\
    \end{pmatrix}, \quad \mu, \mu^\prime > 0,
\end{equation}
where $| \Phi \rangle$ is the vacuum state obtained by solving the RHB equations \cite{RING1984261,VRETENAR2005101}.

We derive the linear response equations by considering the time-dependent generalized density $\mathcal{R}(t)$ in an external charge-changing field $F(t)$ which obeys the equation of motion
\begin{equation}
    i \dot{\mathcal{R}}(t) = [\mathcal{H}(\mathcal{R}(t)) + F(t), \mathcal{R}(t)],
\end{equation}
with $\displaystyle \mathcal{H}=\frac{\partial E}{\partial \mathcal{R}}$.
Assuming harmonic time dependence of the external field
\begin{equation}
    F(t) = Fe^{-i\omega t} + \text{ H.c.},
\end{equation}
with excitation energy $\omega$, and linearizing the generalized density
\begin{equation}
    \mathcal{R}(t) = \mathcal{R}^0 + (\delta \mathcal{R}e^{-i \omega t} + \text{ H.c.}),
\end{equation}
where $[\mathcal{H}(\mathcal{R}^0), \mathcal{R}^0] = 0$ is the static RHB equation, we can derive the Bethe-Salpeter equation for the response $\mathbb{R}$, up to the leading order in $\delta \mathcal{R}$, in the matrix form
\begin{equation}\label{eq:Bethe_salpeter}
    \mathbb{R} = \mathbb{R}^0 + \mathbb{R}^0  \mathbb{W} \mathbb{R},
\end{equation}
defined through the expression 
\begin{equation}
    \delta \mathcal{R}_{\pi \nu} = \sum \limits_{\pi^\prime \nu^\prime} \mathbb{R}_{\pi \nu \pi^\prime \nu^\prime} F_{\pi^\prime \nu^\prime},
\end{equation}
where $\pi$($\nu$) labels proton(neutron) q.p. states. The effective interaction matrix $\mathbb{W}$ is a functional derivative of the interaction Hamiltonian
\begin{equation}
    W_{\pi \nu \pi^\prime \nu^\prime} = \frac{\delta \mathcal{H}_{\pi \nu}}{\delta \mathcal{R}_{\pi^\prime \nu^\prime}},
\end{equation}
while the unperturbed response is diagonal in the q.p. space
\begin{equation}\label{eq:unperturbed_response}
    \mathbb{R}^0_{\pi \nu \pi^\prime \nu^\prime} = \frac{(f_\pi - f_\nu) \delta_{\pi \pi^\prime} \delta_{\nu \nu^\prime
    }}{\omega - E_\pi - E_\nu + i\eta}.
\end{equation}
Small parameter $\eta$ is introduced to avoid the occurrence of singularities and provide finite width of the resonances. In the q.p. basis, matrices $\mathcal{R}^0$ and $\mathcal{H}^0$ are diagonal:
\begin{equation}
    \mathcal{R}^0 = \begin{pmatrix}
        f_\mu & 0 \\
        0 & f_{\tilde{\mu}}
    \end{pmatrix}, \quad \mathcal{H}^0 = \begin{pmatrix}
        E_\mu & 0 \\
        0 & E_{\tilde{\mu}}
    \end{pmatrix},
\end{equation}
with the eigenvalues: 
\begin{align}\label{eq:qp_energies_and_occup}
    \begin{split}
        &f_\mu = 0, \quad f_{\tilde{\mu}} = 1, \\
        &E_\mu = E_k, \quad E_{\tilde{\mu}} = -E_k.
    \end{split}
\end{align}
The formalism can be easily extended to finite-temperature by identifying $f_\mu$ with the Fermi-Dirac factor \cite{PhysRevC.104.064302}. The strength function is defined by contracting the response matrix $\mathbb{R}$ with the external field
\begin{equation}
    S_F(\omega) = - \frac{1}{\pi}\text{Im}\sum \limits_{\pi \nu \pi^\prime \nu^\prime} \left( F_{\pi \nu}^* \mathbb{R}_{\pi \nu \pi^\prime \nu^\prime} F_{\pi^\prime \nu^\prime}\right).
\end{equation}
Although Eq. (\ref{eq:Bethe_salpeter}) can be solved by direct matrix inversion, this is not computationally feasible since the size of the 2 q.p. space increases rapidly with the basis dimension. This problem can be circumvented if the full interaction Hamiltonian can be written as a sum of separable terms
\begin{equation}\label{eq:separable_terms}
    \hat{H} = \hat{H}_0 + \sum \limits_{c c^\prime} v_{c c^\prime} \hat{Q}_c^\dag \hat{Q}_{c^\prime},
\end{equation}
where $\hat{H}_0$ is the mean-field Hamiltonian, and indices $c, c^\prime$ run over a set of operators $\hat{Q}_c$ with coupling strength $v_{c c^\prime}$. In general, index $c$ runs over states in the discretized basis, e.g. coordinate space basis $(r,z)$, momentum space basis $(p_r, p_z)$ or harmonic oscillator basis $(n_r, n_z)$. The interaction matrix $\mathbb{W}$ has the following form \cite{PhysRevC.104.064302}
\begin{equation}\label{eq:interaction_matrix}
    \mathbb{W}_{\pi \nu \pi^\prime \nu^\prime} = \sum \limits_{c c^\prime} v_{c c^\prime}  (Q^{c}_{\pi \nu})^* Q^{c^\prime}_{\pi^\prime \nu^\prime} + v_{c c^\prime}  (Q^c_{\tilde{\pi}^\prime \tilde{\nu}^\prime})^* Q_{\tilde{\pi} \tilde{\nu}}^{c^\prime},
\end{equation}
and we can define the reduced response function
\begin{equation}
\label{eq:reduced_response}
    R_{c c^\prime}(\omega) = \sum \limits_{\pi \nu \pi^\prime \nu^\prime} (Q^c_{\pi \nu})^* \mathbb{R}_{\pi \nu \pi^\prime \nu^\prime} Q_{\pi^\prime \nu^\prime}^{c^\prime},
\end{equation}
as well as the unperturbed reduced response function
\begin{equation}\label{eq:unperturbed_reduced_response}
    R_{c c^\prime}^0(\omega) = \sum \limits_{\pi \nu \pi^\prime \nu^\prime} (Q^c_{\pi \nu})^* \mathbb{R}^0_{\pi \nu \pi^\prime \nu^\prime} Q_{\pi^\prime \nu^\prime}^{c^\prime}.
\end{equation}
The dimension of the $R_{c c^\prime}(\omega)$ and $R_{c c^\prime}^0(\omega)$ matrices is defined
 by the number of interaction channels $N_c$. By plugging the definitions (\ref{eq:interaction_matrix}), (\ref{eq:reduced_response}) and (\ref{eq:unperturbed_reduced_response}) into Eq. (\ref{eq:Bethe_salpeter}), we obtain the following equation
\begin{equation}
\label{eq:R_cF}
    R_{cF}(\omega) = R_{cF}^0(\omega) + \sum \limits_{c^\prime c^{\prime \prime}} R^0_{c c^\prime}(\omega) v_{c^\prime c^{\prime \prime}} R_{c^{\prime \prime} F},
\end{equation}
 where
\begin{equation}
    R_{cF} = \sum \limits_{\pi \nu \pi^\prime \nu^\prime} (Q^c_{\pi \nu})^* \mathbb{R}_{\pi \nu \pi^\prime \nu^\prime} F_{\pi^\prime \nu^\prime}.
\end{equation}
Eq. (\ref{eq:R_cF}) can be solved by inverting the $(1 - R^0 v)$ matrix in the reduced space spanned by interaction channels.
Next, we obtain the response $R_{FF}$ as
\begin{equation}
    R_{FF} = R_{FF}^0 + \sum \limits_{c c^\prime} R^0_{F c} v_{c c^\prime} R_{c^\prime F},
\end{equation}
with the following definitions
\begin{align}
    R_{FF} &= \sum \limits_{\pi \nu \pi^\prime \nu^\prime} \left( F_{\pi \nu}^* \mathbb{R}_{\pi \nu \pi^\prime \nu^\prime} F_{\pi^\prime \nu^\prime}\right), \\
    R_{FF}^0 &= \sum \limits_{\pi \nu \pi^\prime \nu^\prime} \left( F_{\pi \nu}^* \mathbb{R}^0_{\pi \nu \pi^\prime \nu^\prime} F_{\pi^\prime \nu^\prime}\right), \\
    R_{Fc}^0 &= \sum \limits_{\pi \nu \pi^\prime \nu^\prime} \left( F_{\pi \nu}^* \mathbb{R}^0_{\pi \nu \pi^\prime \nu^\prime} Q^c_{\pi^\prime \nu^\prime}\right). 
\end{align}
Finally, the strength function is given by 
\begin{equation}\label{eq:strength_function_Im}
S_F(\omega) = - \frac{1}{\pi} \text{Im} R_{FF}.
\end{equation}

Specific details of above computations can be found in Appendix \ref{sec:appa}.
There are several important differences between the spherical linear response pnRQRPA introduced in Ref. \cite{PhysRevC.104.064302} and the present work. First, in the axial geometry, the pnRQRPA equations cannot be separated into $J^\pi$ blocks, defined by the total angular momentum $J$ and parity $\pi$. However, since the $J_z$ component of the angular momentum operator and the parity operator still commute with the nuclear Hamiltonian, the pnRQRPA equations can still be cast into block diagonal form with $K^\pi$ blocks \cite{PhysRevC.77.044313}. Here, $K$ denotes the eigenvalue of the $J_z$ operator and $\pi$ denotes the parity.

Therefore, the following selection rules are applied when constructing the q.p. pairs within the axially deformed pnRQRPA
\begin{equation}\label{eq:AXDEF_condition}
K = \Omega_p - \Omega_n, \quad \pi = \pi_p \times \pi_n.
\end{equation}
Of course, the dimension of the axially deformed pnRQRPA equation is much larger than the spherical pnRQRPA equation. We notice that in Eq. (\ref{eq:AXDEF_condition}) both the states $k$ with $\Omega_k>0$ and time-reversed states $\bar{k}$ with $\Omega_k < 0$ have to be taken into account, which complicates the expressions for matrix elements. Finally, the number of interaction channels $N_c$ in spherical geometry is determined as a product of the number of terms in the residual interaction and the number of discretized points in the radial mesh. For axially-symmetric geometry, $N_c$ is roughly doubled because we have to discretize the two-dimensional $(r,z)$-mesh.

The single-particle Dirac wave-functions in the coordinate-space have the form \cite{PhysRevC.83.044303}
\begin{equation}\label{eq:wavefunction}
\Psi_k = \frac{1}{\sqrt{2 \pi}} \begin{pmatrix}
f_k^+(r,z) e^{i(\Omega_k - 1/2)\phi} \\
f_k^-(r,z) e^{i(\Omega_k + 1/2)\phi} \\
ig_k^+(r,z) e^{i(\Omega_k - 1/2)\phi} \\
ig_k^-(r,z) e^{i(\Omega_k + 1/2)\phi} \\
\end{pmatrix},
\end{equation}
with the corresponding time-reversed states
\begin{equation}\label{eq:wavefunction_TR}
\Psi_{\bar{k}} = \frac{1}{\sqrt{2 \pi}} \begin{pmatrix}
f_k^-(r,z) e^{-i(\Omega_k + 1/2)\phi} \\
-f_k^+(r,z) e^{-i(\Omega_k - 1/2)\phi} \\
-ig_k^-(r,z) e^{-i(\Omega_k + 1/2)\phi} \\
ig_k^+(r,z) e^{-i(\Omega_k - 1/2)\phi} \\
\end{pmatrix},
\end{equation}
where $f_k^\pm$ are upper and $g_k^\pm$ lower radial components.

In the following, we separately discuss how to calculate the matrix elements of the external field operator, \textit{particle-hole} ($ph$) residual interaction, and the \textit{particle-particle} ($pp$) residual interaction. Finally, we transform from the single-particle to the q.p. basis and make a connection with the linear response formalism of Ref. \cite{PhysRevC.104.064302}.

\subsection{External field matrix elements}
In this work we consider Fermi $(J^\pi = 0^+)$, Gamow-Teller $(J^\pi = 1^+)$, and spin-dipole ($J^\pi = 0^-, 1^-, 2^-$) external field operators. For the Fermi transitions, the only possible mode is $K=0$, while for the Gamow-Teller  and spin-dipole transitions we have to take into account modes $K=0,\pm 1$ and $K=0,\pm 1, \pm 2$, respectively. We notice that modes $K=\pm 1$ and $K=\pm 2$ are degenerate, i.e., it is sufficient to calculate the $K=+1$ and $K=+2$ only.

The external field matrix element is defined as
\begin{equation}
\langle p | F_{JK} | n \rangle =  \int r dr dz d\phi [\Psi_p^\dag F_{JK} \Psi_n].
\end{equation}
We have to consider four combinations of pairs in the previous equation: $pn, p\bar{n}, \bar{p}n$ and $\bar{p}\bar{n}$, where $\bar{n}$ ($\bar{p}$) denotes time-reversed neutron (proton) state. The Fermi transition operator is defined as
\begin{equation}\label{eq:F_F}
    F_{00} = \tau_{\pm},
\end{equation}
where $\tau_{\pm}$ is the isospin raising or lowering operator.  The GT external field operator is defined as
\begin{equation}\label{eq:F_GT}
    F_{1K} = \sigma_{1K} \tau_\pm,
\end{equation}
where $\sigma$ denotes the Pauli spin matrix. The SD operator reads
\begin{equation}\label{eq:F_SD}
    F_{JK} = r\left[\sigma_1 \otimes Y_1\right]_{J K},
\end{equation}
where $Y_{JM}$ is the spherical harmonic and $J = 0,1,2$. The matrix elements of these operators can be readily evaluated in the proton-neutron single-particle basis by using the wavefunctions defined in Eqs. (\ref{eq:wavefunction}) and (\ref{eq:wavefunction_TR}).

Finally, the matrix elements have to be transformed to the q.p. space. This is analogous to Ref. \cite{PhysRevC.104.064302} for spherical pnRQRPA, with the difference that no angular momentum coupling $J$ is performed. In the q.p. basis the external field operator assumes the form
\begin{align}\label{eq:AXDEF_external_field_qp_basis}
\begin{split}
\hat{F}_{JK} &= \sum \limits_{pn} \langle p | F_{JK} | n \rangle c_p^\dag c_n \\
&= \sum \limits_{\pi \nu}  (U^\dag F_{JK} U)_{\pi \nu} \beta_\pi^\dag \beta_\nu + (U^\dag F_{JK} V^*)_{\pi \nu} \beta_\pi^\dag \beta_\nu^\dag \\
&+ (V^T F_{JK} U)_{\pi \nu} \beta_\pi \beta_\nu + (V^T F_{JK} V^*)_{\pi \nu} \beta_\pi \beta_\nu^\dag .\\
\end{split}
\end{align}
Introducing the doubled q.p. basis as in Eq. (\ref{eq:doubled_basis}) we can rewrite above expression as
\begin{align}\label{eq:AXDEF_external_field_qp_basis}
\begin{split}
\hat{F}_{JK} &= \sum \limits_{\pi \nu}  (U^\dag F_{JK} U)_{\pi \nu} a_{\tilde{\pi}} a_\nu + (U^\dag F_{JK} V^*)_{\pi \nu} a_{\tilde{\pi}} a_{\tilde{\nu}} \\
&+ (V^T F_{JK} U)_{\pi \nu} a_\pi a_\nu + (V^T F_{JK} V^*)_{\pi \nu} a_\pi a_{\tilde{\nu}} ,\\
\end{split}
\end{align}
where $\pi, \nu > 0$. Using the matrix structure defined in Eq. (\ref{eq:signature_2qp}), and extending $\pi, \nu$ to negative values we can conveniently rewrite the external field operator as
\begin{equation}
    \hat{F}_{JK} = \sum \limits_{\pi \nu} \begin{pmatrix}
        (U^\dag F_{JK} U)_{\pi \nu} &  (U^\dag F_{JK} V^*)_{\pi \nu} \\
        (V^T F_{JK} U)_{\pi \nu} & (V^T F_{JK} V^*)_{\pi \nu} \\
    \end{pmatrix} a_\pi^\dag a_\nu,
\end{equation}
consistent with derivations in Ref. \cite{PhysRevC.104.064302}.

\begin{table*}
\centering
\caption{Separable matrix elements of the isovector-vector (TV) interaction in the coordinate-space representation [cf. Eq. (\ref{eq:AXDEF_separable_TV})]. We show the matrix elements for the $pn$ and $\bar{p}\bar{n}$ types of transitions.}\label{tab:matrix_elements_TV}
\begin{tabular}{c|c|c}
\hline
\hline
         & $pn$ & $\bar{p} \bar{n}$ \\
         \hline
$Q^{TV(t)}_{pn}(r,z)$ & $f_p^+ f_n^+ + f_p^- f_n^- + g_p^+ g_n^+ + g_p^- g_n^-$ & $f_p^+ f_n^+ + f_p^- f_n^- + g_p^+ g_n^+ + g_p^- g_n^-$\\
$Q^{TV(s),+1}_{pn}(r,z)$ & $i\sqrt{2} [g_p^+ f_n^- - f_p^+ g_n^-]$ & $(+)i\sqrt{2}[g_p^- f_n^+ - f_p^- g_n^+]$\\
$Q^{TV(s),0}_{pn}(r,z)$ & $-i[g_p^+ f_n^+ - g_p^- f_n^- - f_p^+ g_n^+ + f_p^- g_n^-]$ & $-i[g_p^+ f_n^+ - g_p^- f_n^- - f_p^+ g_n^+ + f_p^- g_n^-]$\\
$Q^{TV(s),-1}_{pn}(r,z)$ & $-i\sqrt{2}[g_p^- f_n^+ - f_p^- g_n^+]$ & $(-)i\sqrt{2} [g_p^+ f_n^- - f_p^+ g_n^-]$\\
\hline
\end{tabular}
\end{table*}

\subsection{Particle-hole matrix elements}
In this work we employ two relativistic point-coupling EDFs, DD-PC1 and DD-PCX, which follow the separable form of Eq. (\ref{eq:separable_terms}). Due to the isospin selection rules, only two terms of the interaction Lagrangian density can contribute to the charge-exchange linear response equations. First is the isovector-vector (TV) term with the matrix element
\begin{align}
\begin{split}
V^{TV}_{p n n^\prime p^\prime} &= - \int d^3 \boldsymbol{r}_1 d^3 \boldsymbol{r}_2 \alpha_{TV} [\rho_v] \left[\bar{\Psi}_p(\boldsymbol{r}_1) \gamma_\mu^{(1)} \vec{\tau}^{(1)} \Psi_n(\boldsymbol{r}_1) \right] \times \\
&\times
\left[\bar{\Psi}_{n^\prime}(\boldsymbol{r}_2) \gamma^{\mu (2)} \vec{\tau}^{(2)} \Psi_{p^\prime}(\boldsymbol{r}_2) \right]\delta(\boldsymbol{r}_1 - \boldsymbol{r}_2),
\end{split}
\end{align}
where $\bar{\Psi} = \Psi^\dag \gamma_0$. The coupling $\alpha_{TV}$ is a function of the vector density
\begin{equation}
    \rho_v(\boldsymbol{r}) = \sum \limits_k V_k^\dag(\boldsymbol{r}) V_k(\boldsymbol{r}),
\end{equation}
where $V_k(\boldsymbol{r})$ is the coordinate-space representation of the lower component of the q.p. wavefunction and the summation is performed by omitting the anti-particle states, within the \textit{no-sea} approximation \cite{NIKSIC2011519,GAMBHIR1990132}. We note that the $\alpha_{TV}$ coupling is fully constrained at the ground-state level. The TV residual interaction term can be separated into time-like and space-like components
\begin{widetext}
\begin{align}
\begin{split}
V^{TV(t)}_{p n n^\prime p^\prime} &= -2 \int r dr dz d\phi  \alpha_{TV}[\rho_v] \left[ \Psi_p^\dag(\boldsymbol{r}) \Psi_n(\boldsymbol{r}) \right] \left[ \Psi_{n^\prime}^\dag(\boldsymbol{r}) \Psi_{p^\prime}(\boldsymbol{r}) \right], \\
V^{TV(s)}_{p n n^\prime p^\prime} &= -2 \int r dr dz d\phi  \alpha_{TV}[\rho_v] \sum \limits_\mu (-)^\mu  \left[ \Psi_p^\dag(\boldsymbol{r}) \begin{pmatrix}
0 & \sigma_\mu \\
\sigma_\mu & 0
\end{pmatrix} \Psi_n(\boldsymbol{r}) \right]  \left[ \Psi_{n^\prime}^\dag(\boldsymbol{r}) \begin{pmatrix}
0 & \sigma_{-\mu} \\
\sigma_{-\mu} & 0
\end{pmatrix} \Psi_{p^\prime}(\boldsymbol{r}) \right], \\
\end{split}
\end{align}
\end{widetext}
where factor 2 originates from the isospin matrix element. We see that interaction can be written in a separable form, where the separable channels are defined as
\begin{align}\label{eq:AXDEF_separable_TV}
&Q_{pn}^{TV(t)}(r,z) = \Psi_p^\dag(r,z) \Psi_n(r,z), \\
 &Q^{TV(s),\mu}_{pn} = \Psi_p^\dag(r,z) \begin{pmatrix}
0 & \sigma_\mu \\
\sigma_\mu & 0
\end{pmatrix} \Psi_n(r,z).
\end{align}
The integration over the $\phi$ angle can be performed analytically and provides the selection rules for the angular momentum projections. The total number of the separable channels for the TV interaction term is $4 \times N_z^{GH} \times N_r^{GL}$, where $N_z^{GH}$ and $N_r^{GL}$ are the number of Gauss-Hermite and Gauss-Laguerre integration mesh-points in the $z$- and $r$-directions, respectively. In table \ref{tab:matrix_elements_TV} we show the separable channels of the TV interaction for $pn$ and $\bar{p}\bar{n}$ types of the transitions in the coordinate-space basis.

\begin{table*}
\centering
\caption{Same as in table \ref{tab:matrix_elements_TV} but for the TPV interaction.}\label{tab:matrix_elements_TPV}
\begin{tabular}{c|c|c}
\hline
\hline
         & $pn$ & $\bar{p} \bar{n}$ \\
         \hline
$Q^{TPV(t)}_{pn}(r,z)$ & $i[f_p^+ g_n^+ + f_p^- g_n^- - g_p^+ f_n^+ - g_p^- g_n^-]$ & $(-i)[f_p^+ g_n^+ + f_p^- g_n^- - g_p^+ f_n^+ - g_p^- f_n^-]$\\
$Q^{TPV(s),+1}_{pn}(r,z)$ & $
- \sqrt{2}[f_p^+ f_n^- + g_p^+ g_n^-]$ & $\sqrt{2}[f_p^- f_n^+ + g_p^- g_n^+]$\\
$Q^{TPV(s),0}_{pn}(r,z)$ & $f_p^+ f_n^+ + g_p^+ g_n^+ - f_p^- f_n^- - g_p^- g_n^-$ & $(-)[f_p^+ f_n^+ + g_p^+ g_n^+ - f_p^- f_n^- - g_p^- g_n^-]$\\
$Q^{TPV(s),-1}_{pn}(r,z)$ & $\sqrt{2}[f_p^- f_n^+ + g_p^- g_n^+]$ & $-\sqrt{2}[f_p^+ f_n^- + g_p^+ g_n^-]$\\
\hline
\end{tabular}
\end{table*}

Although the pion does not contribute at the Hartree level, it can have significant impact on the pnRQRPA residual interaction \cite{PhysRevC.69.054303,PhysRevC.103.064307}. Therefore, we also include the isovector-pseudovector (TPV) term in the residual interaction
\begin{align}
\begin{split}
V^{TPV}_{p n n^\prime p^\prime} &= g_0 \int d^3 \boldsymbol{r}_1 d^3 \boldsymbol{r}_2 \left[\bar{\Psi}_p(\boldsymbol{r}_1) \gamma_5^{(1)} \gamma_\mu^{(1)} \vec{\tau}^{(1)} \Psi_n(\boldsymbol{r}_1) \right] \times \\
&\times
\left[\bar{\Psi}_{n^\prime}(\boldsymbol{r}_2) \gamma_5^{(2)} \gamma^{\mu (2)} \vec{\tau}^{(2)} \Psi_{p^\prime}(\boldsymbol{r}_2) \right]\delta(\boldsymbol{r}_1 - \boldsymbol{r}_2).
\end{split}
\end{align}
We note that the Landau-Migdal strength parameter $g_0$ is unconstrained at the ground-state level.
Therefore, in this work we use $g_0 = 0.734$(0.621) for the DD-PC1 (DD-PCX) EDFs, adjusted to reproduce the experimental GT centroid in ${}^{208}$Pb (see Refs. \cite{PhysRevC.104.064302,PhysRevC.103.064307}).

Analogously to the TV case, the isovector-pseudovector (TPV) residual interaction can be written as
\begin{widetext}
\begin{align}
\begin{split}
V^{TPV(t)}_{p n n^\prime p^\prime} &= 2 g_0 \int r dr dz d\phi   \left[ \Psi_p^\dag(\boldsymbol{r}) \begin{pmatrix}
0 & 1 \\
1 & 0
\end{pmatrix} \Psi_n(\boldsymbol{r}) \right] \left[ \Psi_{n^\prime}^\dag(\boldsymbol{r}) \begin{pmatrix}
0 & 1 \\
1 & 0
\end{pmatrix} \Psi_{p^\prime}(\boldsymbol{r}) \right], \\
V^{TPV(s)}_{p n n^\prime p^\prime} &= 2 g_0 \int r dr dz d\phi   \sum \limits_\mu (-)^\mu  \left[ \Psi_p^\dag(\boldsymbol{r}) \begin{pmatrix}
\sigma_\mu & 0 \\
0 & \sigma_\mu
\end{pmatrix} \Psi_n(\boldsymbol{r}) \right] \left[ \Psi_{n^\prime}^\dag(\boldsymbol{r}) \begin{pmatrix}
\sigma_{-\mu} & 0 \\
0 & \sigma_{-\mu}
\end{pmatrix} \Psi_{p^\prime}(\boldsymbol{r}) \right], \\
\end{split}
\end{align}
\end{widetext}
for time-like and space-like components, respectively. The separable channels are defined as
\begin{align}\label{eq:AXDEF_separable_TPV}
&Q_{pn}^{TPV(t)}(r,z) =  \Psi_p^\dag(r,z) \begin{pmatrix}
0 & 1 \\
1 & 0
\end{pmatrix} \Psi_n(r,z), \\
&Q^{TPV(s),\mu}_{pn} = \Psi_p^\dag(r,z) \begin{pmatrix}
\sigma_\mu & 0 \\
0 & \sigma_\mu
\end{pmatrix} \Psi_n(r,z).
\end{align}
Therefore, the total dimension of the TPV separable channels is also $4 \times N_{z}^{GH} \times N_{r}^{GL}$. The corresponding matrix elements in the coordinate-space basis are shown in table \ref{tab:matrix_elements_TPV}.

The separable matrix elements are transformed to the q.p. basis analogously to the external field matrix elements in Eq. (\ref{eq:AXDEF_external_field_qp_basis})
\begin{align}\label{eq:AXDEF_ph_field_qp_basis}
\begin{split}
\hat{Q}_{c c^\prime} &= \sum \limits_{\pi \nu} \begin{pmatrix}
(U^\dag Q_{c c^\prime} U)_{\pi \nu} & (U^\dag Q_{c c^\prime} V^*)_{\pi \nu} \\
(V^T Q_{c c^\prime} U)_{\pi \nu} & (V^T Q_{c c^\prime} V^*)_{\pi \nu}
\end{pmatrix} a_\pi^\dag a_\nu,
\end{split}
\end{align}
where $(c, c^\prime)$ label the separable interaction channels.

\subsection{Particle-particle matrix elements}\label{sec:AXDEF_pp}

For the particle-particle ($pp$) interaction we assume the separable pairing form \cite{PhysRevC.80.024313}

\begin{equation}
V^\prime(\boldsymbol{r}_1, \boldsymbol{r}_2, \boldsymbol{r}_1^\prime, \boldsymbol{r}_2^\prime) = -f G\delta(\boldsymbol{R} - \boldsymbol{R}^\prime) P(r,z) P(r^\prime, z^\prime),
\end{equation}
where $\boldsymbol{R} = \frac{1}{2}(\boldsymbol{r}_1 + \boldsymbol{r}_2)$ is the center-of-mass and $\boldsymbol{r} = \boldsymbol{r}_1 - \boldsymbol{r}_2$ is the relative coordinate. The overall factor $f$ is defined as
\begin{equation}
    f = \left\{ \begin{array}{c}
         V_0^{pp}, \quad T = 0, S = 1  \\
         1, \quad T = 1, S = 0 
    \end{array} \right. ,
\end{equation}
where $V_0^{pp}$ is the isoscalar pairing strength, not constrained at the ground-state level \cite{PhysRevC.104.064302}.
The form factor $P(r,z)$ corresponds to the Gaussian function
\begin{equation}
P(r,z) = \frac{1}{(4 \pi a^2)^{3/2}} e^{- \frac{z^2 + r^2}{4a^2}},
\end{equation}
with strength $G$ and range $a$ parameters adjusted to reproduce the pairing gap of the Gogny pairing force \cite{PhysRevC.80.024313}. 
It is convenient to calculate the matrix element in the axially-deformed h.o. basis
\begin{align}
\begin{split}
\langle 1 2 | V | 1^\prime 2^\prime \rangle &= \langle 1 2 | V^\prime(1 - P^r P^\sigma P^\tau)  | 1^\prime 2^\prime \rangle,
\end{split}
\end{align}
where each state is denoted with the h.o. quantum numbers $| 1 \rangle \equiv | n_{z_1} n_{r_1} \Lambda_1 m_{s_1} m_{t_1} \rangle$, where $n_{z}$, and $n_r$ are quantum numbers in $z$, and $r$ directions, respectively. $\Lambda$ is the projection of the orbital angular momentum on the $z$ axis, $m_s$ is the spin projection, and $m_{t}$ denotes the isospin projection. In the coordinate-space the h.o. eigenfunction has the form
\begin{align}
\begin{split}
\langle \boldsymbol{r} | n_{z_1} n_{r_1} \Lambda_1 m_{s_1} m_{t_1} \rangle &= \phi_{n_{z_1}}(z,b_z) \phi_{n_{r_1}}^{\Lambda_1}(r,b_r) \\
&\times \frac{e^{i \phi \Lambda_1}}{\sqrt{2\pi}} \chi_{1/2 m_{s_1}} \xi_{1/2 m_{t_1}},    \end{split}
\end{align}
where $\chi_{1/2 m_{s_1}}$ denotes the spin, and $\xi_{1/2 m_{t_1}}$ the isospin wavefunctions. The $b_r$ and $b_z$ are the oscillator lengths, defined in Ref. \cite{PhysRevC.80.024313}.

The projector operators exchange the position, spin, and isospin of two nucleons
\begin{align}
\begin{split}
&P^r |\boldsymbol{r}_1 \boldsymbol{r}_2 \rangle = | \boldsymbol{r}_2 \boldsymbol{r}_1 \rangle, \quad P^\sigma |S M_S \rangle = (-)^{S-1} |S M_S \rangle, \\
&P^\tau| T M_T \rangle = (-)^{T-1} | T M_T \rangle,
\end{split}
\end{align}
where $S$ and $T$ denote the total spin and isospin of two states, with projections $M_S$ and $M_T$, respectively. The wave function coupled to total spin $S$ and isospin $T$ reads
\begin{align}
\begin{split}
|1 2 \rangle &= \phi_{n_{z_1}}(z_1, b_z) \phi_{n_{r_1}}^{\Lambda_{1}} (r_1, b_r) \phi_{n_{z_2}}(z_2, b_z) \phi_{n_{r_2}}^{\Lambda_{2}} (r_2, b_r) \\
&\times \frac{1}{2\pi} e^{i \phi_1 \Lambda_1} e^{i \phi_2 \Lambda_2}  \sum \limits_{S M_S} C^{S M_S}_{1/2 m_{s_1} 1/2 m_{s_2}} | S M_S \rangle \\
&\times\sum \limits_{T M_T} C^{T M_T}_{1/2 m_{t_1} 1/2 m_{t_2}} | T M_T \rangle.
\end{split}
\end{align}

In order to calculate the matrix elements, we have to transform the h.o. wave functions from the laboratory to the center-of-mass frame. First, the product of $z$-component wave functions can be written as (see Ref. \cite{BJELCIC2020107184} and references therein)
\begin{equation}
\phi_{n_{z_1}}(z_1) \phi_{n_{z_2}}(z_2) = \sum \limits_{N_z n_z} M^{n_{z_1} n_{z_2}}_{N_z n_z} \phi_{N_z}(Z, \tilde{b}_Z) \phi_{n_z}(z, \tilde{b}_z) (-)^{n_z},
\end{equation}
where $\tilde{b}_Z = \sqrt{2}b_z$ and $\tilde{b}_z = b_z/\sqrt{2}${\color{blue}.} $M_{N_z n_z}^{n_{z_1} n_{z_2}}$ is the 1-dimensional Talmi-Moshinsky coefficient \cite{MOSHINSKY1959104}. Next, we apply the same transformation to the radial wave functions \cite{BJELCIC2020107184}
\begin{align}
\begin{split}
\phi_{n_{r_1}}^{\Lambda_1}(\boldsymbol{r}_1) \phi_{n_{r_2}}^{\Lambda_2}(\boldsymbol{r}_2) &= \sum \limits_{N_r \Lambda} \sum \limits_{n_r \lambda} M^{n_{r_1} \Lambda_1 n_{r_2} \Lambda_2}_{N_r \Lambda n_r \lambda}  \\
&\times \phi_{N_r}^\Lambda(\boldsymbol{R}, \tilde{b}_R) \phi_{n_r}^\lambda(\boldsymbol{r}, \tilde{b}_r) (-)^\lambda,
\end{split}
\end{align}
where $\tilde{b}_R = \sqrt{2}b_r$ and $\tilde{b}_r = b_r / \sqrt{2}${\color{blue}.} $M^{n_{r_1} \Lambda_1 n_{r_2} \Lambda_2}_{N_r \Lambda n_r \lambda}$ is the 2-dimensional Talmi-Moshinsky coefficient \cite{MOSHINSKY1959104}. The Talmi-Moshinsky coefficients imply the following selection rule that connects quantum numbers in the intrinsic and laboratory frame \cite{BJELCIC2020107184}
\begin{align}
&n_{z_1} + n_{z_2} = N_z + n_z, \\
&n_r + N_r = n_{r_1} + n_{r_2} + \frac{|\Lambda_1| + |\Lambda_2| + |\Lambda_1 + \Lambda_2| }{2}, \\ 
&\Lambda_1 + \Lambda_2 = \Lambda + \lambda.
\end{align}
The total matrix element in the coupled basis reads
\begin{align}
\begin{split}
\langle 1 \bar{2} | V | 1^\prime \bar{2}^\prime \rangle &= - G\delta_{\lambda 0} \delta_{\lambda^\prime 0} \delta_{\Lambda \Lambda^\prime} \frac{1}{b_z b_r^2} \Sigma(S,T)   \\
&\times \sum \limits_{N_z N_r} W_{1 \bar{2}}^{N_z} W_{1 \bar{2}}^{N_r} W_{1^\prime \bar{2}^\prime}^{N_z} W_{1^\prime \bar{2}^\prime}^{N_z} \\
\end{split}
\end{align}
where we have defined the separable terms analogously to Ref. \cite{NIKSIC20141808}:
\begin{align}
\begin{split}
W_{12}^{N_z} &= \frac{1}{\sqrt{b_z}} M^{N_z n_z}_{n_{z_1} n_{z_2}} \delta_{n_z, \text{even}} \frac{(-)^{n_z/2}}{(2\pi)^{1/4}} \frac{\sqrt{n_z !}}{2^{n_z/2} (n_z/2)!}\\
&\times \left( \frac{b_z^2}{a^2 + b_z^2} \right)^{1/2} \left( \frac{b_z^2 - a^2}{b_z^2 + a^2} \right)^{n_z/2},
\end{split}
\end{align}
\begin{align}
W_{12}^{N_r} &= \frac{1}{b_r} M^{N_r \Lambda n_r 0}_{n_{r_1} \Lambda_1 n_{r_2} \Lambda_2} \frac{1}{(2\pi)^{1/2}} \frac{b_r^2}{b_r^2 + a^2} \left( \frac{b_r^2 - a^2}{b_r^2 + a^2} \right)^{n_r}.
\end{align}
We notice that $n_z$ can only assume even values. Spin-isospin part $\Sigma(S,T)$ has the form
\begin{align}
\begin{split}
&\Sigma(S,T) = \sum \limits_{S M_S} \sum \limits_{T M_T} \frac{1}{2}[1 - (-)^{S+T} ](-)^{1/2 - m_{s_2}} (-)^{1/2 - m_{s_2}^\prime}  \\
&\times C^{S M_S}_{1/2 m_{s_1} 1/2 -m_{s_2}} C^{S M_S}_{1/2 m_{s_1}^\prime 1/2 -m_{s_2}^\prime} \\
& \times C^{T M_T}_{1/2 m_{t_1}^\prime 1/2 m_{t_2}^\prime} C^{T M_T}_{1/2 m_{t_1} 1/2 m_{t_2}},
\end{split}
\end{align}
from which it follows that $S+T$ assumes only odd values. Two cases can be distinguished corresponding to either \textbf{isovector} ($T = 1, S = 0$) or \textbf{isoscalar} ($T = 0, S = 1$) pairing interaction. The separable matrix element for the isovector pairing is characterized by $N_r, N_z$ quantum numbers and has the form
\begin{equation}
W_{N_r, N_z}^{T = 1, S = 0} = \frac{1}{\sqrt{2}} W^{N_z}_{1 \bar{2}} W^{N_r}_{1 \bar{2}} (-)^{1/2 - m_{s_2}} C^{0 0}_{1/2 m_{s_1} 1/2 -m_{s_2}},
\end{equation}
where the $1/\sqrt{2}$ factor stems from the isospin part and $C^{1 0}_{1/2 - 1/2 1/2 +1/2 } = 1/\sqrt{2}$. On the other hand, the isoscalar matrix element is determined by the $M_S$ quantum number, in addition to $N_r, N_z$
\begin{equation}
W_{N_r, N_z, M_S}^{T = 0, S = 1} = -\frac{1}{\sqrt{2}} W^{N_z}_{1 \bar{2}} W^{N_r}_{1 \bar{2}} (-)^{1/2 - m_{s_2}} C^{1 M_S}_{1/2 m_{s_1} 1/2 -m_{s_2}}.
\end{equation}
The total $pp$ residual interaction matrix element can be written in the following form
\begin{align}
\begin{split}
V^{pp}_{p n p^\prime n^\prime} &= \langle p \bar{n} | V | p^\prime \bar{n}^\prime \rangle c_p^\dag c_{\bar{n}}^\dag c_{\bar{n}^\prime} c_{p^\prime}  \\
&= \sum \limits_{N_z N_r M_S} \left( W^{N_z N_r M_S}_{pn} \right)^* c_p^\dag c_{\bar{n}}^\dag  W^{N_z N_r M_S}_{p^\prime n^\prime} c_{\bar{n}^\prime} c_{p^\prime}  \\
&= \sum \limits_{N_z N_r M_S} \left(\hat{Q}^{N_z N_r M_S }_{pn} \right)^\dag \hat{Q}_{p^\prime n^\prime}^{N_z N_r M_S},
\end{split}
\end{align}
where we have defined a separable term as $\left(\hat{Q}^{N_z N_r M_S }_{pn} \right)^\dag =  \left( W^{N_z N_r M_S}_{pn} \right)^*  c_p^\dag c_{\bar{n}}^\dag$. The total number of separable matrix elements for the isovector pairing interaction is $N_r \times N_z$, while for the isoscalar pairing interaction, it is $3 \times N_z \times N_r$ (factor 3 comes from projections of spin $S = 1$). Next, we have to transform the $pp$ separable matrix elements from the single-particle to the q.p. basis. Here one has to take into account the transformation properties of time-reversed states \cite{ring2004nuclear} to obtain the correct expression
\begin{align}\label{eq:AXDEF_pp_field_qp_basis}
\begin{split}
\hat{W}_{c c^\prime} &= \sum \limits_{\pi \nu}  \begin{pmatrix}
-(U^T W_{c c^\prime} V^*)_{\pi \nu} & (U^\dag W_{cc^\prime} U)_{\pi \nu} \\
-(V^T W_{c c^\prime} V^*)_{\pi \nu} & (V^T W_{c c^\prime} U^*)_{\pi \nu}
\end{pmatrix} a_\pi^\dag a_\nu,
\end{split}
\end{align}
where $(c, c^\prime)$ denote the separable $pp$ residual interaction channels.

\section{Numerical tests}\label{sec:numerical_tests}
First numerical test is performed for doubly-magic neutron-rich ${}^{28}$O isotope. To prevent the pairing collapse, we have artificially increased the pairing strength of the DD-PC1 effective interaction from $G_{p,n} = -728$ MeV fm${}^{-3}$ to $G_{p,n} = -1500$ MeV fm${}^{-3}$. This value is sufficient to break the shell closure in ${}^{28}$O thus providing a stringent test for the implementation of the $pp$-channel in the residual interaction. For initial numerical tests, we employ a small basis space of $N_{osc} = 8$ h.o. shells without any additional cut-off on the 2 q.p. basis. 
As a rule of thumb, we have found that reasonable convergence of the integrals can be achieved by using $N_{z}^{GL} = N_r^{GL} \sim N_{osc}$ mesh points. 
Furthermore, the smearing parameter is set to small value $\eta=0.25$ MeV to provide better resolution of individual peaks in the strength function. Results calculated with the axially-deformed pnRQRPA are compared with the spherical pnRQRPA from Ref. \cite{PhysRevC.104.064302}. To make the comparison meaningful, the axially-deformed RHB calculations are constrained to a spherical shape. Throughout this work, excitation energies $\omega$ from Eq. (\ref{eq:strength_function_Im}) are shown with respect to the parent nucleus.

\begin{figure}[t!]
\centering
\includegraphics[width=\linewidth]{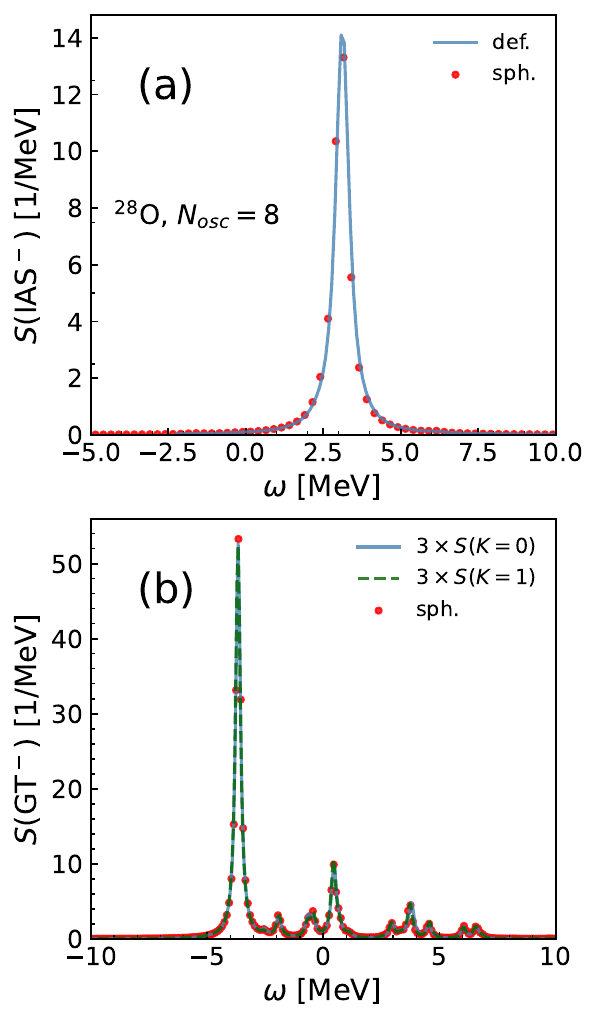}
\caption{Comparison between the spherical and axially-deformed pnRQRPA results for ${}^{28}$O for the IAS${}^-$ (a) and GT${}^{-}$ (b) strength function, with $N_{osc} = 8$ oscillator shells. The strength function calculated with the spherical pnRQRPA is represented with red circles. Different components of the axially-deformed response are also shown: $K = 0$ mode is represented with solid blue and $K = 1$ mode with dashed green. The response function of the deformed pnRQRPA is multiplied by 3 to account for degeneracy.}\label{fig:AXDEF_num_test_1}
\end{figure}

In Fig. \ref{fig:AXDEF_num_test_1}(a) we show the Fermi strength function in the $\beta^-$ direction (IAS${}^-$). The results obtained with deformed and spherical pnRQRPA code are in excellent agreement.  In Fig. \ref{fig:AXDEF_num_test_1}(b) we display the GT${}^-$ strength function for both $K = 0$ and $K = 1$ modes. In spherical limit, all three modes $K=\pm 1$ and $K=0$ should be degenerated and this result is reproduced by our calculation. Due to the degeneracy, the total GT strength is  $S(\text{GT}^-) = 3 \times S(K = 0) = 3 \times S(K = 1)$, where $S(K=0,1)$ denotes the strength function for the $K = 0,1$ mode. We note that the total strength agrees with spherical calculation. We note that the excitation energy with respect to the parent nucleus $\omega$ can be negative if the binding energy of the daughter nucleus is smaller than that of the parent, \textit{i.e.}, ground state of the daughter lies below that of the parent \cite{PhysRevC.101.044305}.

\begin{figure}[t!]
\centering
\includegraphics[width=\linewidth]{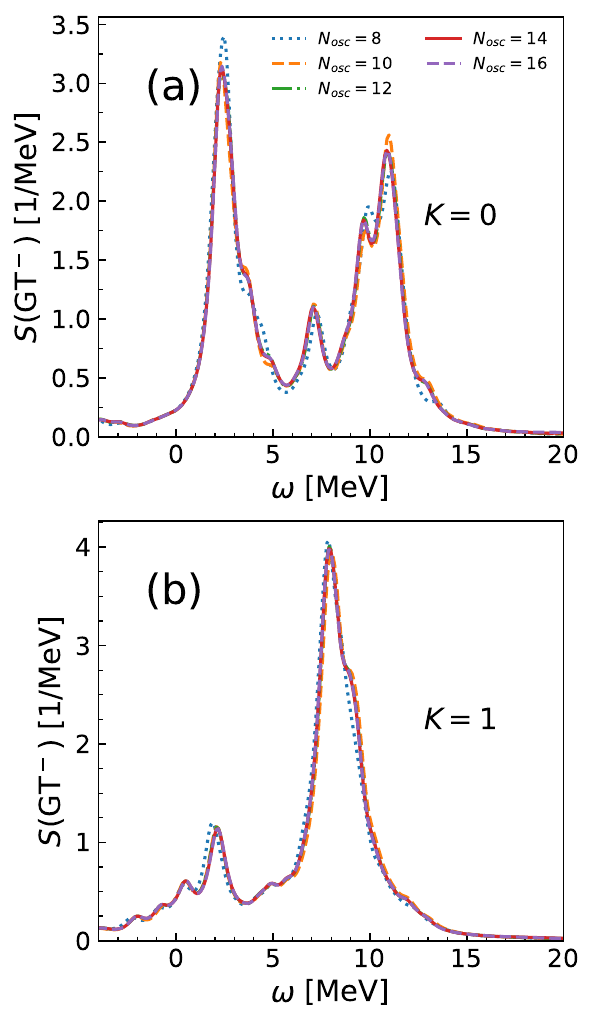}
\caption{Convergence tests of the GT${}^-$ strength for ${}^{70}$Fe with $\beta_2 = +0.3$ for a varying number of oscillator shells $N_{osc}$ and no additional cut-off to the 2 q.p. basis. Results are shown for the $K = 0$ (a) and $K = 1$ (b) projections.}\label{fig:AXDEF_num_test_3}
\end{figure}

Next, we would like to determine the basis size necessary for performing reliable calculations of spin-isospin response in medium mass nuclei. For this purpose, we have calculated the GT$^{-}$ response in neutron-rich $^{70}$Fe isotope for several numbers of h.o. shells ranging from $N_{osc} = 8$ to $N_{osc} = 16$. The RHB ground state in each calculation is constrained to $\beta_2=0.3$.  Results are shown in Fig. \ref{fig:AXDEF_num_test_3} for both $K = 0$ (a) and $K = 1$ (b) projections. From the figure, we conclude that reasonable convergence for the GT${}^-$ strength is achieved already with $N_{osc} = 12$. Furthermore, the strength functions for $N_{osc} = 14$ and $N_{osc} = 16$ are almost indistinguishable. We note that this conclusion agrees with calculations based on non-relativistic EDF presented in 
Ref. \cite{PhysRevC.94.014304}, where good convergence for $A \sim 70$ nuclei is obtained with $N_{osc} = 13$.

Therefore, in the following calculations, we use $N_{osc} = 16$ which guarantees good convergence for $pf$-shell nuclei considered in this work. We notice that the number of 2 q.p. pairs for $N_{osc} = 16$ becomes very large. In table \ref{tab:AXDEF_num_of_pairs} we show the total number of 2 q.p. pairs $N_{pair}$ for GT$^{-}$ ($K = 0$ and $K = 1$) modes in ${}^{70}$Fe isotope with $\beta_2 = +0.3$. Therefore, solving for one projection with $N_{osc} = 16$ would require diagonalization of a square matrix with dimension $180000$. Here, the linear response formalism based on the reduced response function for separable interaction is advantageous. We have to perform a sum over $N_{pair}$ and invert a matrix with moderate size of $5400 \times 5400$. We notice that the sum over the 2 q.p. pairs can be easily parallelized thus reducing the computation time.

\begin{table}
\centering
\caption{Number of proton-neutron 2 q.p. pairs $N_{pair}$ in $^{70}$Fe for $K = 0$ and $K = 1$ projections of the Gamow-Teller response for an increasing number of oscillator shells $N_{osc}$.}\label{tab:AXDEF_num_of_pairs}
\begin{tabular}{ccc}
\hline
\hline
$N_{osc}$ & $N_{pair}(K = 0)$ & $N_{pair}(K = 1)$ \\
\hline
8 & 5002 & 4857 \\
10 & 12444 & 12188 \\
12 & 26894 & 26481 \\
14 & 52432 & 51808 \\
16 & 94482 & 93585 \\
\hline 
\end{tabular}
\end{table}

Finally, by performing the singular-value decomposition (SVD) of the $Q^c$ matrices, one can reduce the number of interaction channels thus speeding-up matrix-multiplication operations while preserving the accuracy. Details  of this procedure are discussed in Appendix~\ref{sec:appb}.

\section{Spin-isospin excitations in medium heavy axially-deformed nuclei}
\label{sec:AXDEF_spin_isospin}

In this section we study the effects of deformations on the spin-isospin response in selected medium heavy nuclei. Calculations are performed with the DD-PC1 \cite{PhysRevC.78.034318} and DD-PCX \cite{PhysRevC.99.034318} relativistic EDFs, $N_{osc} = 16$ h.o. shells and no other truncation on the 2 q.p. basis is imposed. The contributions from the anti-particle transitions are neglected because their influence on the charge-exchange excitations are negligible \cite{PhysRevC.107.014318}. The strength functions are smeared with $\eta = 1$ MeV. For unnatural parity transitions $J^\pi = 0^-, 1^+,$ and $2^-$, the isoscalar pairing strength is set to $V_0^{pp} = 1.0$. Since the main aim of this work is to study the influence of deformation on spin-isospin strength functions, detailed optimization of 
$V_0^{pp}$ is left for future work, and we refer the reader to Refs. \cite{PhysRevC.105.064315,PhysRevC.93.025805}.

\begin{figure*}[ht!]
\centering
\includegraphics[width=\linewidth]{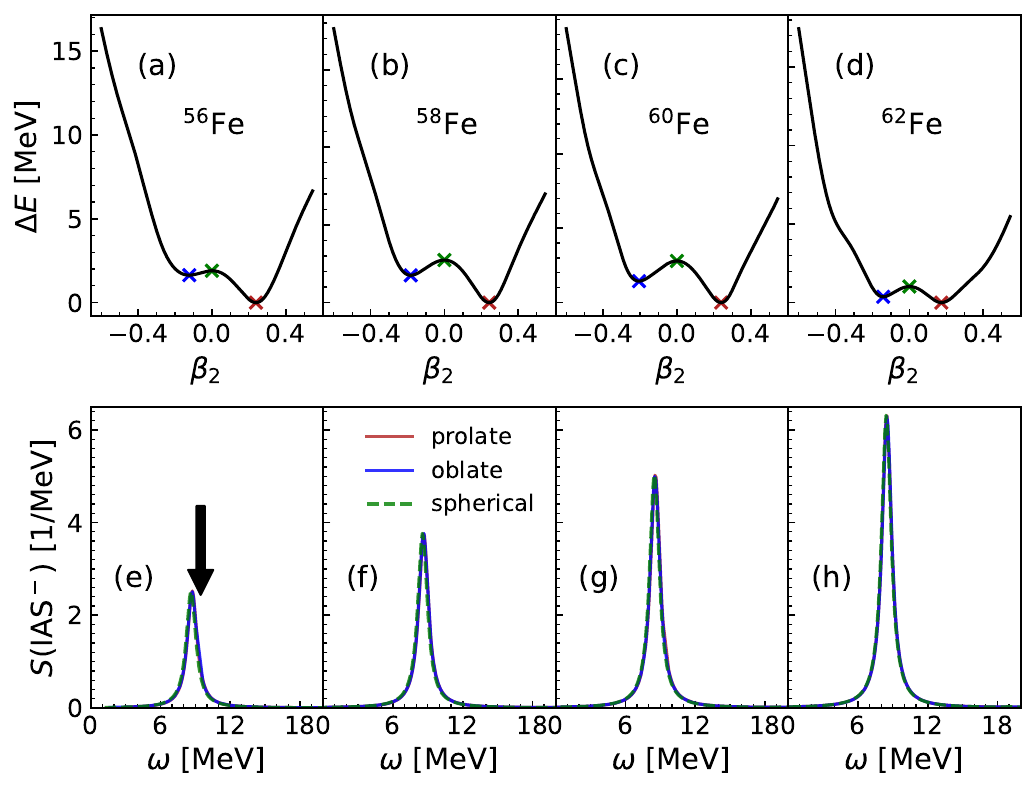}
\caption{Panels (a)--(d): potential energy curves for $^{56-62}$Fe isotopes as functions of the $\beta_2$ deformation, calculated with the DD-PC1 interaction. For each isotope, the energy is normalized with respect to the binding energy of the global minimum. The crosses mark locations of oblate, spherical, and prolate configurations used in the subsequent pnRQRPA calculations. Panels (e)--(h): the IAS${}^{-}$ strength function for $^{56-62}$Fe isotopes calculated on top of the oblate, spherical, and prolate configurations marked with crosses in upper panels. The experimental centroid energy from Ref. \cite{PhysRevC.88.054329} is denoted with a black arrow.}\label{fig:AXDEF_IAS_Fe_chain}
\end{figure*}
\subsection{The Isobaric Analog Resonance}

In the following we study the Fermi $(J^\pi = 0^+)$ strength function in the $^{56-62}$Fe isotopes. The potential energy curves (PEC), calculated with the DD-PC1 interactions, are displayed in the upper panels (a)-(d) of Fig. \ref{fig:AXDEF_IAS_Fe_chain}. For each isotope, the energy is normalized with respect to the binding energy of the global minimum. The pnRQRPA calculations are performed on top of the configurations marked with crosses that correspond to oblate, spherical and prolate shapes.

Simple structure of the Fermi operator, as in Eq. (\ref{eq:F_F}), allows only transitions with the same quantum numbers in the Nilsson basis. Furthermore, only the $K = 0$ component of the angular momentum projection is allowed. One of the important characteristics of the Isobaric Analog Resonance (IAR) is its narrow width. This is because it has the same isospin as the parent state, while the neighboring states have the isospin of the ground state of the daughter nucleus, i.e.,
they differ in isospin by one unit. This means that they will
couple only weakly with the IAR. The excitation energy of IAR corresponds to the difference between the even-even parent and odd-odd daughter nucleus Coulomb energy, corrected by the residual interaction. 

In lower panels of Fig. \ref{fig:AXDEF_IAS_Fe_chain}(e)--(h) we show the Fermi strength function for ${}^{56,58,60,62}$Fe for prolate, oblate, and spherical configurations. We observe that deformation has almost no influence on the Fermi strength function, i.e. the differences between the position of IAR for prolate, oblate and spherical configurations do not exceed $0.05$ MeV.  This is indeed an expected result because small quadrupole deformations induce only second-order effect on the total Coulomb energy ($\sim \beta_2^2$) \cite{ring2004nuclear}. The experimental centroid energy, obtained from the (${}^{3}$He,$t$) charge-exchange reaction in Ref. \cite{PhysRevC.88.054329}, is denoted by a black arrow. The strong IAR was extracted at $\omega \approx 8.9$ MeV, around $0.5$ MeV higher in comparison to our calculations. We have also performed calculations with the DD-PCX interaction for the ${}^{56}$Fe isotope and found that the strength is shifted around 0.13 MeV to higher energies, slightly closer to the experimental data. Although we have found that the Fermi strength function is almost independent of the deformation, calculations presented in this section still provide a reliable test of our numerical implementation of the deformed pnRQRPA. 

\subsection{The Gamow-Teller resonance}\label{sec:AXDEF_GT}

Next we study the response to the GT operator defined in Eq. (\ref{eq:F_GT}). The Pauli spin matrix in the GT operator induces the selection rules for spin and angular momentum $\Delta S = 1$ and $\Delta L = 0$. The calculations require strength functions for two projections $K = 0$ and 
$K = 1$ because of the degeneracy of $K=+1$ and $K=-1$ modes. The total strength is calculated as
\begin{equation}\label{eq:AXDEF_total_strength}
S(\text{GT}^\pm,\omega) = S(K = 0, \omega) + 2 \times S(K = 1, \omega).
\end{equation}
Because of the selection rule for $pp$ matrix elements $S+T=\textnormal{odd}$, only the isoscalar pairing $(S=1,T=0)$ contributes to the $pp$ residual interaction.

\begin{figure*}[t!]
\centering
\includegraphics[scale=0.8]{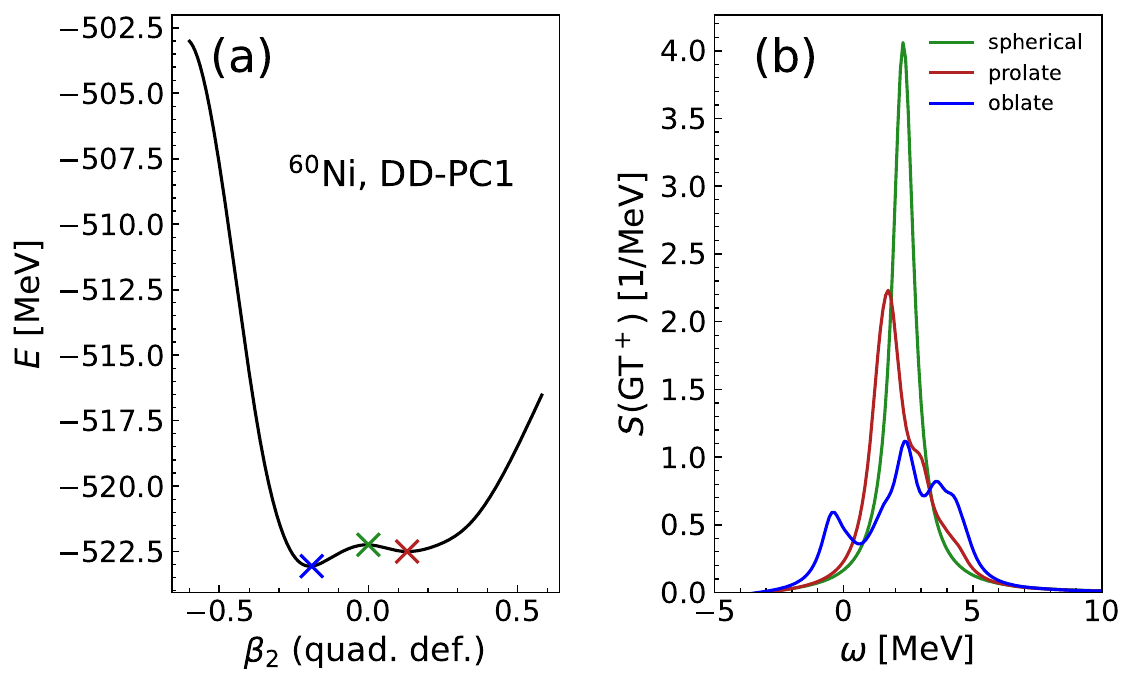}
\caption{(a) The PEC for ${}^{60}$Ni calculated with the axially-deformed RHB with the DD-PC1 interaction. Three stationary points  (marked with crosses) correspond to the oblate (blue), spherical (green), and prolate (red) configurations. (b) The GT${}^+$ strength as a function of the excitation energy $\omega$. Calculations are performed on top of the three selected configurations marked with crosses in panel (a).}\label{fig:AXDEF_GT_60Ni}
\end{figure*}

As a first example we study the role of deformation effects on the GT${}^+$ strength in ${}^{60}$Ni isotope. First we perform a constrained RHB calculation and the resulting potential energy curve is displayed in Fig. \ref{fig:AXDEF_GT_60Ni} (a). We select three points on the PEC: oblate minimum at $\beta_2 = -0.19$, local maximum at spherical point, and prolate local minimum $\beta_2 = 0.13$. On top of these three configurations we have performed the pnRQRPA calculation for the GT${}^+$ response. The resulting strength is displayed in Fig. \ref{fig:AXDEF_GT_60Ni} (b) as a function of the excitation energy. We notice that the oblate configuration shows pronounced fragmentation of the strength function. On the other hand, for the prolate shape, the fragmentation is reduced. However, the broadening of the resonance around $\omega \sim 3$ MeV suggests the existence of the second peak. This is a consequence of using $\eta = 1$ MeV as the smearing width, effectively masking the strength. In fact, we have verified that by using a smaller value $\eta = 0.5$ MeV instead, a clear separation of the main peak at $\omega \sim 1.7$ MeV and another peak at $\omega \sim 3$ MeV is observed. The fragmentation is more pronounced for the oblate configuration because of the larger value of the quadrupole moment in comparison to the prolate configuration, causing deformation to play a more significant role. On the other hand, strength function for the spherical configuration is concentrated in a single peak at $\omega = 2.3$ MeV. Of course, the fragmentation in the deformed 
pnRQRPA strength stems from the degeneracy breaking of the 
Nilsson q.p. orbitals.

\begin{figure*}[ht!]
\centering
\includegraphics[width=\linewidth]{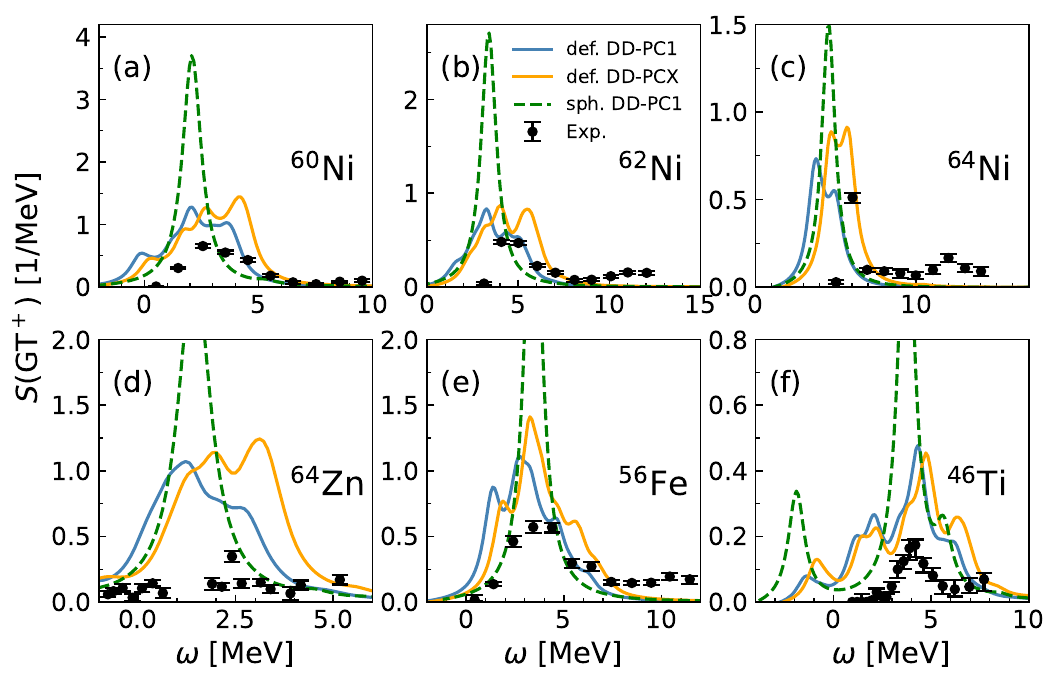}
\caption{The GT${}^+$ strength function for selected $pf$-shell nuclei as a function of the excitation energy $\omega$. The calculations are performed with the axially-deformed pnRQRPA by employing the DD-PC1 (blue solid line) and DD-PCX (orange solid line) interactions and the spherical pnRQRPA for the DD-PC1 interaction (green dashed line). The results are compared with the available experimental data from Refs. \cite{PhysRevC.51.1144,PhysRevLett.112.252501,PhysRevC.80.014313,PhysRevC.49.3128} (black circles). The isoscalar pairing strength is set to $V^{pp}_{0} = 1.0$ in all calculations.}\label{fig:AXDEF_GT_strength_comparison_with_exp}
\end{figure*}

It is instructive to compare the GT${}^+$ strength function with the available experimental data. Unfortunately, the GT strength function has been measured only for a handful of nuclei in the $pf$-shell and a limited range of excitation energies \cite{PhysRevC.49.3128,PhysRevC.51.1144,PhysRevLett.112.252501,PhysRevC.80.014313}. This means that only a part of the total strength function is accessible in the experiment and often it is not the part with the main resonance peak.  

In Fig. \ref{fig:AXDEF_GT_strength_comparison_with_exp} panels (a)-(f) we compare the GT${}^+$ strength function in $^{60,62,64}$Ni, $^{64}$Zn, $^{56}$Fe and $^{46}$Ti with available experimental data. Calculations were performed by employing DD-PC1 (blue solid line) and DD-PCX (orange solid line) effective interactions.  First, we note that all presented nuclei display axially-deformed shapes in the ground state, either prolate (${}^{56}$Fe and ${}^{46}$Ti) or oblate (${}^{60,62,64}$Ni and ${}^{64}$Zn). We have summarized the values of the ground state quadrupole deformations $\beta_2$ for selected nuclei in Tab. \ref{tab:AXDEF_optimal_defs}. In Fig. \ref{fig:AXDEF_GT_strength_comparison_with_exp} we have also included the 
GT${}^+$ strength function calculated with the DD-PC1 effective interaction on top of the spherical configurations for each isotope (green dashed line).

\begin{table}
\centering
\caption{The optimal quadrupole deformation $\beta_2$ for selected $pf$-shell nuclei using both the DD-PC1 and DD-PCX interactions.}\label{tab:AXDEF_optimal_defs}
\begin{tabular}{ccc}
\hline
\hline
nucleus & $\beta_2$(DD-PC1) & $\beta_2$(DD-PCX) \\
\hline
${}^{60}$Ni &  -0.19 & - 0.16 \\
${}^{62}$Ni & -0.22 & -0.18 \\
${}^{64}$Ni & -0.13 & -0.09 \\
${}^{64}$Zn & -0.24 & -0.13 \\
${}^{56}$Fe & 0.24 & 0.21 \\
${}^{46}$Ti & 0.24 & 0.22 \\
\hline
\end{tabular}
\end{table}

\begin{figure*}[ht!]
\centering
\includegraphics[width=\linewidth]{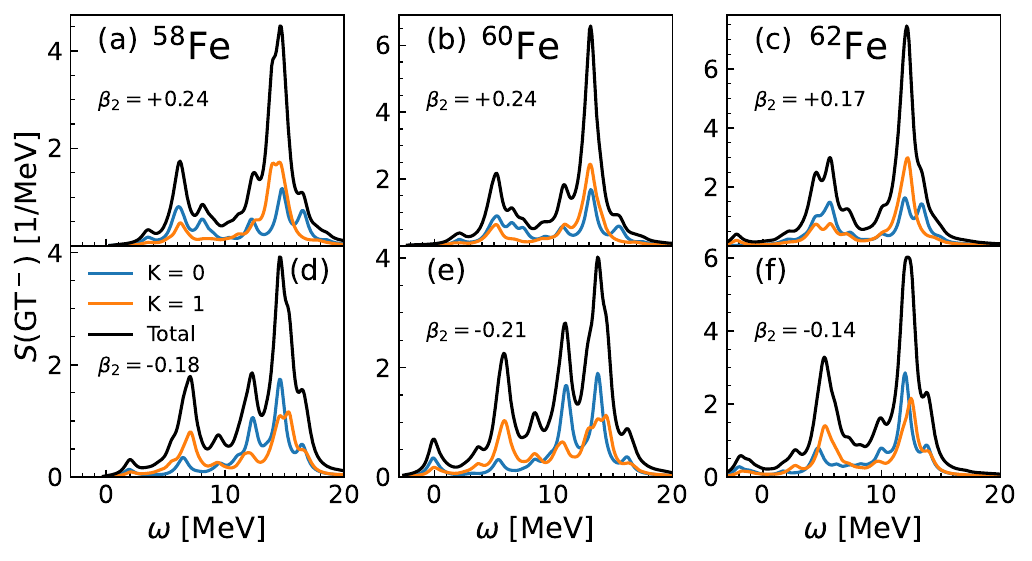}
\caption{The GT${}^{-}$ strength function for selected even-even isotopes of iron, shown for the prolate (a)-(c), and oblate (d)-(f) configuration. The total strength function (solid black) is decomposed to the $K = 0$ (solid blue) and $K = 1$ (solid orange) projections of the total angular momentum $J = 1$.}\label{fig:AXDEF_GT_Fe_chain_detail}
\end{figure*}

From Fig. \ref{fig:AXDEF_GT_strength_comparison_with_exp} one can observe a large discrepancy between spherical and deformed calculations. The spherical strength functions (green dashed line) are concentrated in a one resonance peak for all nuclei except for ${}^{46}$Ti that displays more structure in the spherical GT${}^+$ response. On the other hand, the strength functions calculated on top of the deformed configurations display pronounced fragmentation with reduced strength. Difference between spherical and deformed calculations can be attributed to increased density of states and splittings between the Nilsson orbitals for deformed configurations.  
Overall, the calculated strength functions based on deformed configurations are in better agreement with the experiment. For instance, deformation effects lead to very good agreement with experiment for oblate deformed ${}^{60}$Ni and ${}^{62}$Ni isotopes. For ${}^{64}$Ni, the deformation effects lead to splitting of the main resonance peak with reduced strength, in better agreement with the experimental data. For ${}^{64}$Zn we notice that calculations significantly overestimate the measured strength, which predicts no noticeable resonance structure. The deformed pnRQRPA predicts fragmentation of the main resonance peak around $\omega \sim 1$ MeV. In prolate deformed ${}^{56}$Fe and ${}^{46}$Ti, the inclusion of deformation effects improves the agreement with the experiment. However, we notice that our calculation predicts some peaks that are not present in experimental data, especially at lower excitation energies. Since the deformation effects are included, we expect that the differences between the experiment and our calculations can be attributed mainly to coupling with higher-order configurations, as well as the configuration mixing. Namely, the relatively simple pnRQRPA theory includes only the contribution of 2 q.p. excitations to the response function. Expanding the present formalism by including the coupling of 2 q.p. excitations to the phonons (QPVC) would lead to more fragmentation of the strength function, and possibly a better agreement with the experimental data \cite{Robin2016,LITVINOVA2020135134,NIU2018325,PhysRevLett.114.142501}. However, we note that the deformed QPVC at the level of the residual interaction is at present in its nascent phase \cite{PhysRevC.109.044308,PhysRevC.105.044326}, due to significant numerical challenges. In addition, it is known that optimization of isoscalar pairing strength $V_0^{pp}$ can lead to better agreement of strength centroids \cite{PhysRevC.105.064309,PhysRevC.69.054303}, however, we leave these efforts for future work. Furthermore, Eq. (\ref{eq:AXDEF_total_strength}) is only an approximation valid for large deformations (so-called \textit{needle approximation}). The problem is that the transformation from the intrinsic system of the nucleus to the laboratory system has to be performed, which mixes contributions of different angular momenta $J$. Therefore, a proper projection method for the response of the deformed nuclei should be implemented as discussed in Refs. \cite{Zeh1967,PhysRevC.77.034317}.


To assess the possible systematic uncertainties within our calculations, we also perform the deformed pnRQRPA calculations by employing the DD-PCX interaction (solid orange line). The DD-PCX relativistic EDF was adjusted to both the ground-state properties and excitations in atomic nuclei \cite{PhysRevC.99.034318}. From Fig. \ref{fig:AXDEF_GT_strength_comparison_with_exp} we observe significant differences in the strength function calculated with two interactions. For instance, the agreement with experimental data deteriorates slightly for ${}^{60,62}$Ni isotopes, as the strength function for the DD-PCX interaction is pushed to slightly higher excitation energies. On the other hand, although the strength is overestimated, the DD-PCX leads to better agreement with the experimental centroid in the ${}^{64}$Ni isotope. For other nuclei we also observe a slight shift of the strength centroid to higher excitation energies. For ${}^{56}$Fe, the strength function calculated with the DD-PCX interaction is more concentrated around the main resonance peak, thus better describing the experimental distribution. In the case of ${}^{46}$Ti, the strength functions calculated with the DD-PCX and DD-PC1 interactions have similar shape, although the DD-PCX strength function is shifted to higher energy around $\sim 0.5$ MeV. On the other hand, for ${}^{64}$Zn this shift is more pronounced being around 1.5 MeV, hence the experimental centroid is better reproduced with the DD-PCX interaction. Therefore, based on our calculations with two different functionals we can conclude that the GT strength function is very sensitive not only to deformation of the atomic nucleus, but also to details of the effective interaction used in the calculations. Furthermore, including the deformation effects in the calculation considerably improves the overall agreement between the theoretical and experimental GT${}^+$ strength function.

\begin{figure*}[ht!]
\centering
\includegraphics[width=\linewidth]{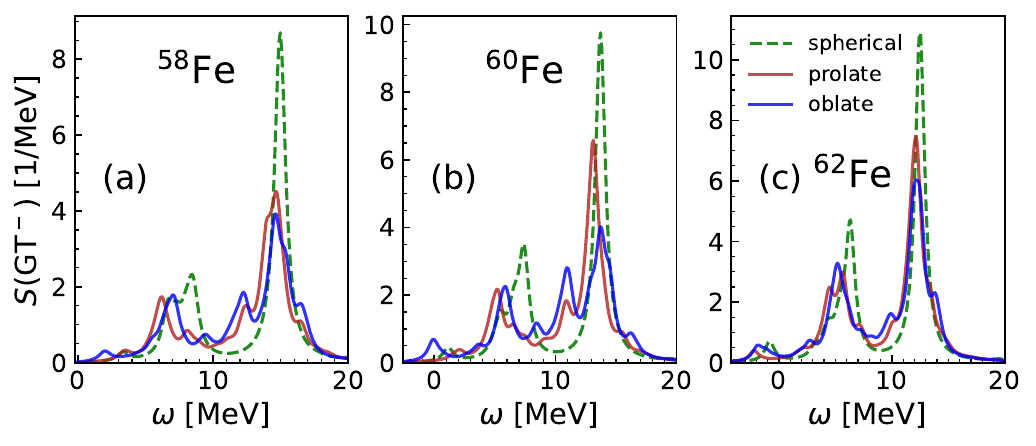}
\caption{The total GT${}^{-}$ strength function in ${}^{58}$Fe, ${}^{60}$Fe and ${}^{62}$Fe, shown for the prolate (solid red), oblate (solid blue) and spherical (dashed green) configuration.}\label{fig:AXDEF_GT_Fe_chain}
\end{figure*}

In the following, we turn our attention to the GT${}^-$ strength function. Nuclei displaying significant GT${}^-$ strength are often neutron-rich, thus obtaining the experimental data is much more difficult. In fact, most of the experimental data exist for nuclei around the shell closure, such as tin isotopes \cite{PhysRevC.51.526}. Due to the proximity of shell closure, these nuclei are most often spherical, therefore considering the deformation effects is of no importance for the strength function. We have already compared the results of our spherical pnRQRPA with the DD-PC1 interaction for particular tin isotopes in Ref. \cite{PhysRevC.104.064302} with available experimental data, and obtained excellent agreement of the strength centroids. The low-lying GT${}^-$ strength function is especially important for calculating the $\beta$-decay half-lives, since part of the low-lying strength is contained within the $Q_\beta$ energy window. We investigate the GT${}^-$ strength function for ${}^{58,60,62}$Fe, with the deformed pnRQRPA using the DD-PC1 interaction. All three nuclei display oblate and prolate minima in the potential energy curve, therefore, we can study the influence of deformation on the GT${}^-$ strength function. In Fig. \ref{fig:AXDEF_GT_Fe_chain_detail}(a)-(f) we show the GT${}^-$ strength function for selected iron isotopes. Solid blue and orange lines denote $K=0$ and $K=1$ components, while solid black line denotes the total strength calculated by using Eq. (\ref{eq:AXDEF_total_strength}). Calculations based on prolate minima are displayed in Fig.~\ref{fig:AXDEF_GT_Fe_chain_detail} panels (a)-(c), while those based on oblate minima are displated in Fig.~\ref{fig:AXDEF_GT_Fe_chain_detail} panels (d)-(f). We observe that the GT${}^-$ strength function consists of the low-lying peak and a resonance peak (GTR) located at higher excitation energies. A similar structure was also obtained in spherical calculations \cite{PhysRevC.104.064302}. In comparison to results presented in Ref. \cite{PhysRevC.104.064302}, the deformed GT${}^-$ response function displays more complicated structure. It is interesting to notice that for prolate shapes ($\beta_2 > 0$), the dominant contribution to the low-lying strength comes from the $K=0$ component, while the strength in resonance region is dominated by the $K = 1$ component. On the other hand, the opposite is true for the oblate shape ($\beta_2 < 0$). In spherical nuclei, both $K = 0$ and $K = 1$ modes are degenerate, however, in deformed nuclei, the degeneracy is broken and these modes split. For the oblate configurations, the $K = 0$ mode is pushed towards the lower excitation energies and $K = 1$ towards higher. The opposite is true for the prolate configurations. The amount of splitting between the modes is proportional to the magnitude of $\beta_2$. We notice that similar degeneracy splitting was already observed in Refs. \cite{PhysRevC.77.034317,PhysRevC.78.064316,PhysRevC.83.021304} for the like-particle response function and in Refs. \cite{Yoshida_ptep,PhysRevC.90.024308} for the charge-exchange case.

Finally, in Fig. \ref{fig:AXDEF_GT_Fe_chain}(a)-(c) we show the total GT${}^-$ strength function for the prolate, oblate, and spherical configurations. Compared to the spherical strength function, which consists mainly of two peaks, the deformed strength function displays more complicated structure. For nuclei with larger quadrupole deformations, ${}^{58}$Fe and ${}^{60}$Fe, we observe a larger difference compared to the spherical strength function. For these nuclei, oblate configurations show more fragmentation in the GTR region in comparison to the prolate ones. By inspecting Fig. \ref{fig:AXDEF_GT_Fe_chain_detail}(d)-(f) we observe that a large splitting of the GTR strength originates from the $K = 0$ mode, which is more dominant at higher excitation energies. On the other hand, ${}^{62}$Fe has a lower value of $\beta_2$ compared to the ${}^{58,60}$Fe isotopes, thus the differences between strength functions for spherical, prolate and oblate configurations are reduced. Overall, a significantly richer structure predicted by deformed pnRQRPA follows from a higher density of states for axially-deformed nuclei compared to the spherical ones. Therefore, we expect that deformed calculations will predict more strength contributing to the $Q_\beta$ window and therefore they could significantly alter $\beta$-decay half-lives compared to spherical calculations.

\begin{figure*}
    \centering
    \includegraphics[width=\linewidth]{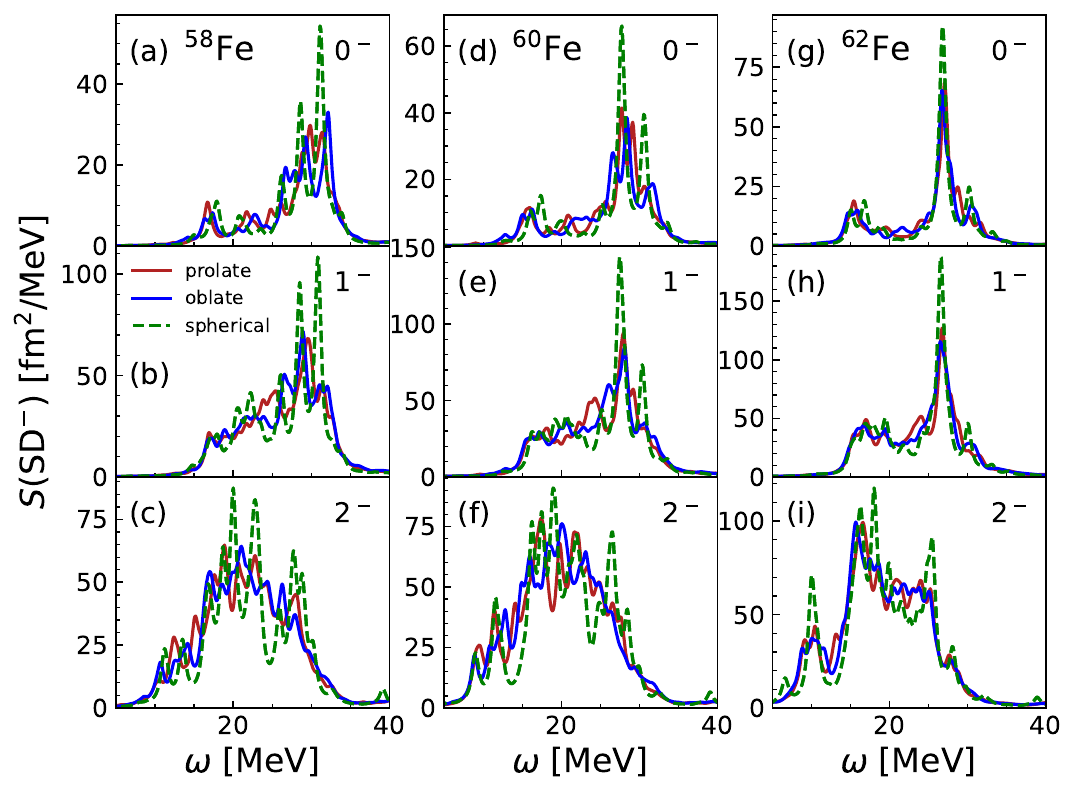}
    \caption{The SD${}^-$ strength function for prolate (solid red), oblate (solid blue), and spherical (dashed green) configurations in ${}^{58}$Fe (a)--(c), ${}^{60}$Fe (d)--(f), and ${}^{62}$Fe (g)--(i). The total SD${}^-$ contribution is split into $0^-$ (upper panels), $1^-$ (middle panels) and $2^-$ (lower panels).}\label{fig:SD}
\end{figure*}

\subsection{The Spin-Dipole Resonance}
The spin-dipole operator was introduced in Eq. (\ref{eq:F_SD}). It corresponds to transitions coupled to total spin $\Delta S = 1$ and orbital angular momentum $\Delta L = 1$. This results in coupling to three possible values of angular momenta, $J^\pi = 0^-, 1^-$, and $2^-$. The energy non-weighted  moment of the SD strength $m_0$ is proportional to the difference between neutron and proton root-mean-square radii \cite{PhysRevLett.82.3216}. Therefore, measurements of the SD transition strength can provide
constraint for the nuclear equation of state (EOS) parameters through extracting the neutron skin thickness \cite{ROCAMAZA201896}. Most of the theoretical and experimental effort is focused on the GT and Fermi transitions. On the other hand, results for the SD are scarce, and mostly limited to spherical nuclei \cite{PhysRevLett.101.122502,Colo2014,PhysRevC.102.054336,PhysRevLett.82.3216}. While for the $0^-$ transition we have to calculate only the $K = 0$ mode, the expression for the total strength function of $1^-$ mode is similar to that of the GT in Eq. (\ref{eq:AXDEF_total_strength}). The strength for $2^-$ mode, assuming time-reversal symmetry, is given by
\begin{align}
\begin{split}
    S(\text{SD}^-,\omega) &= S(K = 0, \omega) + 2 \times S(K=1, \omega) \\ &+ 2 \times S(K=2,\omega).
\end{split}
\end{align}

\begin{figure*}
    \centering
    \includegraphics[width=\linewidth]{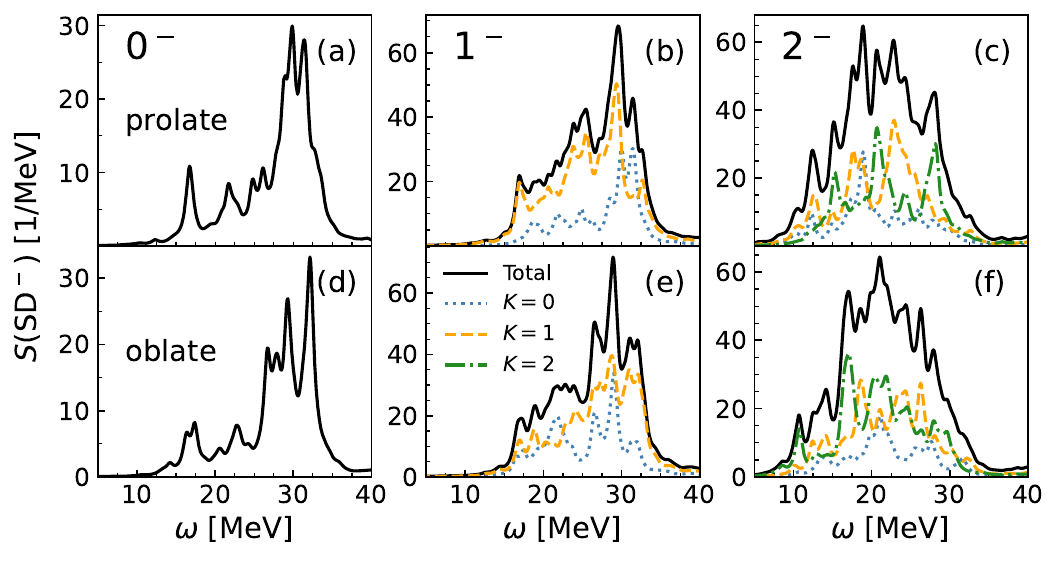}
    \caption{The SD${}^-$ strength function for ${}^{58}$Fe decomposed to different $K$ modes for prolate (a)--(c) and oblate (d)--(f) configurations of $0^-, 1^-$, and $2^-$ multipoles.}
    \label{fig:sdrm_detail}
\end{figure*}

Unlike Fermi and GT operators, SD operator introduces the radial dependence in the matrix elements which leads to an overall richer structure of the transition strength, since the overlap between basis states with different radial dependence can be non-vanishing. The first-forbidden (FF) weak-interaction rates, e.g. in the $\beta$-decay, have a large contribution from the spin-dipole operator \cite{BEHRENS1971111}, and it is well-known that FF transitions can play a large role in determining the total rates \cite{PhysRevC.93.014304,PhysRevC.93.025805}. Therefore, it is important to study the influence of the deformation, not only on the GT, but also on the SD strength function.

\begin{table*}[t!]
    \centering
        \caption{The centroid energy $E_{cent.}$ (in MeV) of spin-dipole $0^-, 1^-$, and $2^-$ excitations for selected iron isotopes and oblate, spherical, or prolate constrained shapes.}\label{tab:centroids}
    \begin{tabular}{c|ccc|ccc|ccc}
    \hline
    \hline
    &     & ${}^{58}$Fe & &  & ${}^{60}$Fe & & & ${}^{62}$Fe & \\
    &  oblate    & spherical & prolate & oblate  & spherical & prolate & oblate & spherical & prolate \\
    \hline
$0^-$ & 28.333 & 28.445 & 28.368 & 26.407 & 26.856 & 26.538 & 25.098 & 25.398 & 25.133 \\
$1^-$ & 27.074 & 26.855 & 26.972 & 25.203 & 25.414 & 25.194 & 24.135 & 24.089 & 24.112 \\
$2^-$ & 22.039 & 22.374 & 21.970 & 20.268 & 20.759 & 20.269 & 19.117 & 19.351 & 19.111 \\
\hline
    \end{tabular}
    \label{tab:my_label}
\end{table*}

In Fig. \ref{fig:SD} we display the total SD${}^-$ strength function in ${}^{58,60,62}$Fe isotopes for $J^\pi = 0^-, 1^-$, and $2^-$ modes. Calculations are performed with the DD-PC1 interaction for three selected configurations in each isotope: oblate, spherical and prolate. Deformed configurations correspond to local minima on the potential energy curve with $\beta_2$ values denoted in Fig. \ref{fig:AXDEF_IAS_Fe_chain}, while spherical configuration corresponds to the local maximum on the potential energy curve. Furthermore, in table \ref{tab:centroids} we show the location of centroids $E_{cent.}$, for each multipole and different nuclear shapes, defined as $E_{cent.} = m_1 / m_0$, where $m_k = \int d \omega \omega^k S(\text{SD}^-, \omega)$ is $k$-th moment of the strength distribution.
First of all, as discussed in Ref. \cite{PhysRevC.104.064302}, we notice a clear separation of centroid energies $E_{cent.}$ for different $J^\pi$ modes within the same nucleus, irrespective of its shape. In particular, as can be observed in table \ref{tab:centroids}, the ordering $E_{cent.}(2^-) < E_{cent.}(1^-) < E_{cent.}(0^-)$ reflects the fact that higher-rank operators are usually easier to excite since the multipole selection rules allow for more configurations to enter into the strength function \cite{PhysRevLett.101.122502,Colo2014,PhysRevC.102.054336}.
Overall, the strength functions are rather complicated and for deformed shapes   more fragmented in comparison to the spherical shape.

We can also get a grasp of what happens as the neutron number is increased. First of all, as can be seen in table \ref{tab:centroids}, there is a systematic shift of the strength function towards lower excitation energies, and the overall strength function increases, since SD modes become easier to excite in neutron-rich nuclei. Due to lowering of $|\beta_2|$ for ${}^{62}$Fe, we observe a formation of a stronger resonance peaks for $0^-,1^-$ transitions around $\omega \sim 30$ MeV. On the other hand, the $2^-$ transition strength is broadly distributed, although the strength function appears less fragmented compared to ${}^{58}$Fe and ${}^{60}$Fe.

To study the mechanism driving the shape evolution of the SD${}^-$ strength within Fe isotopes, we focus on ${}^{58}$Fe and in Fig. \ref{fig:sdrm_detail} plot the SD${}^-$ strength for $J^\pi=0^-$ [panels (a) and (d)], $1^-$ [panels (b) and (e)], and $2^-$ [panels (c) and (f)] multipoles. Calculations are performed with DD-PC1 interaction for prolate [panels (a), (b) and (c)] and oblate [panels (d), (e) and (f)] configurations. The total strength is denoted by solid black line, while the $K=0$, $K=1$ and $K=2$ components are denoted by dotted blue, dashed orange and dashed-dotted green lines, respectively. 
We start with the $0^-$ transitions 
for which only the $K = 0$ mode is possible. The resonance region is found around $\omega \sim 30$ MeV and displays more fragmentation for oblate configurations, similar to the GT case. The $1^-$ strength has more complicated structure in comparison to the GT strength function. We can observe the splitting between the $K = 0$ and $K = 1$ modes depending on the nuclear shape. For the prolate configuration, the $K = 0$ centroid is located at 27.5 MeV, while the $K = 1$ centroid is at 25.7 MeV. On the other hand, for oblate configuration, the $K = 0$ strength is pushed downwards in energy with centroid at 25.5 MeV, while the $K = 1$ strength centroid is pushed upward to 26.9 MeV. Although the conclusions are the same as for GT strength, the mechanism responsible for the splitting of the SD $1^{-}$ strength is more involved (see discussion in Ref. \cite{PhysRevC.102.054336}). For the $2^-$ mode, we also have a contribution from the $K = 2$ projection. For prolate shapes, strength around $\omega \sim 30$ MeV is dominated by the $K = 2$ component, while for the oblate shape, $K = 2$ component determines the low-lying strength around $\omega \sim 10$ MeV. 

From the experimental perspective, one cannot disentangle different $K$ modes, or multipoles, from the SD strength function, therefore, it remains a significant challenge to study shape-induced effects on the SD strength.

\section{Conclusions}
In this work, we have developed an axially-deformed pnRQRPA based on covariant energy density functional theory. The solver utilizes point-coupling EDFs with separable pairing interaction for which the residual interaction Hamiltonian can be written as a sum of a products of separable terms. Since the dimension of the two quasiparticle space increases rapidly with the number of oscillator shells, it is advantageous to reformulate the Bethe-Salpeter equation in terms of the reduced response functions represented by separable interaction channels, thus considerably reducing the computational cost and allowing for large-scale calculations. 

By employing the deformed pnRQRPA, we have investigated the impact of the deformation of atomic nucleus on the strength functions for several types of spin-isospin excitations. We have found that the deformation effects do not contribute to the Fermi transitions because the quadrupole deformation leads to second-order correction to the Coulomb energy which determines the position of the IAS centroid. For the GT strength, we show that deformation effects lead to splitting between the $K = 0$ and $K = 1$ modes, resulting in pronounced fragmentation of the strength function. In fact, by comparing our results with the experimental GT${}^+$ strength, a systematic improvement over the spherical pnRQRPA is achieved. The impact of different interactions is investigated by employing two different functionals, DD-PC1 and DD-PCX. Although we have found that the strength function is sensitive to the details of the effective interaction, our main conclusions related to the deformation effects remain valid. The impact of the nuclear geometry is studied by constraining the nuclear shape to spherical, oblate, and prolate configurations, which leads to considerable changes in the strength function. In particular, the direction of the $K = 0$ and $K = 1$ splitting in deformed nuclei is proportional to the strength of quadrupole deformation $|\beta_2|$, while its direction depends on the sign of $\beta_2$. Finally, the SD transitions show a complicated pattern. The fragmentation of strength for deformed shapes is still pronounced, especially for $0^-$ and $1^-$ modes. Due to the less restrictive selection rule for $2^-$ transitions, already the spherical strength shows a complicated structure without clear resonance peaks, which is again more fragmented for deformed nuclei.
Therefore, the deformation of atomic nucleus has a pronounced effects on the spin-isospin resonances. The development of efficient QRPA solvers together with a considerable increase in computing power in recent years, finally make large-scale calculations for the response of axially-deformed nuclei feasible. Thus the pnRQRPA introduced in this work also open perspectives for the future studies of deformation effects on astrophysically relevant weak interaction processes in the relativistic EDF framework, e.g., beta decay \cite{RavlicPRC.104.054318} and electron capture \cite{RavlicPRC.102.065804} and their role in supernova evolution \cite{GiraudPRC.105.055801}.

The formalism developed in this work could easily be extended to finite temperature, which would enable a study of competition between nuclear shape, pairing, and temperature effects. Of course, one has to be careful when applying pnRQRPA to global calculations across the nuclide chart. First, without any angular momentum projections, results apply to well-deformed nuclei. Second, coupling to higher-order correlations going beyond 2 q.p. excitations, requires the development of axially-deformed QPVC. For the latter, the formalism developed in this work presents a suitable starting point for coupling to like-particle phonons. These developments will be addressed in future work.

\begin{acknowledgments}
This work is supported by the US National Science Foundation under Grant PHY-1927130 (AccelNet-WOU: International Research Network for Nuclear Astrophysics [IReNA]), the National Key Research and Development (R\&D) Program of China under Grant No. 2021YFA1601500 and Natural Science Foundation of China under Grant No.12075104, Croatian Science Foundation under the project number IP-2022-10-7773 and by the project “Implementation of cutting-edge research and its application as part of the Scientific Center of Excellence for Quantum and Complex Systems, and Representations of Lie Algebras“, PK.1.1.02, European Union, European Regional Development Fund, and by the Deutsche Forschungsgemeinschaft (DFG, German Research Foundation)
under Germanys Excellence Strategy EXC-2094-390783311, ORIGINS. Helpful discussions with Gabriel Mart\'{i}nez-Pinedo, Caroline Robin, Thomas Neff, Antonio Bjel\v{c}i\'c, Diana A. Terrero, and Luis Gonz\'alez-Miret are gratefully acknowledged.
\end{acknowledgments}

\appendix

\section{Calculating the reduced response function $R_{c c^\prime}$}\label{sec:appa}
 In this appendix we present some additional details of the linear response equations in axial geometry. We assume $\pi, \nu > 0$ [cf. Eq. (\ref{eq:doubled_basis})] and explicitly label each q.p. component with $j = 1,\ldots,4$. Numerically most intensive part of the calculation is to construct the unperturbed reduced response defined in Eq. (\ref{eq:unperturbed_reduced_response}). This equation can be recast into the matrix form as
\begin{equation}\label{eq:AXDEF_r0_matrix}
R^0(\omega) = \sum_{j=1,4} Q^T_j  N(\omega)_j  Q_j,
\end{equation}
where $R^0(\omega) \in \mathbb{C}^{N_c \times N_c}$, $Q_j \in \mathbb{R}^{N_{pair} \times N_c}$, and $N(\omega)_j \in \mathbb{C}^{N_{pair} \times N_{pair}}$. 
The total number of interaction channels $N_c$ can be written as a sum of the number of $ph$ and $pp$ channels, $N_{c} = N_{ph} + N_{pp}$. $N_{pair}$ denotes the total number of q.p. pairs.
The $Q_j$ matrix includes separable channels of the residual interaction Hamiltonian and has the following form
\begin{equation}
Q_j = \begin{pmatrix}
Q_{i_1}^{1,j} & \ldots & Q_{i_1}^{N_{ph},j}  & W_{i_1}^{1,j} & \ldots &  W_{i_1}^{N_{pp},j} \\
\vdots      &   \ddots        &      \vdots                     &    \vdots                &   \ddots         &          \vdots \\
Q_{i_{N_{pair}}}^{1,j} & \ldots & Q_{i_{N_{pair}}}^{N_{ph},j}  & W_{i_{N_{pair}}}^{1,j} & \ldots &  W_{i_{N_{pair}}}^{N_{pp},j} \\
\end{pmatrix},
\end{equation}
where first superscript denotes the interaction channel (from $1$ to $N_{ph}$ and $N_{pp}$ for $ph$ and $pp$ interaction channels respectively), second superscript ($j=1,\dots,4$) denotes combinations of the q.p. components and subscript denotes the two q.p. pairs from $i_1 \equiv (\pi_1, \nu_1)$ to $i_{N_{pair}} \equiv (\pi_{N_{pair}}, \nu_{N_{pair}})$. For instance, the separable $ph$ matrix elements for the pair $i_1$ will have the following form
\begin{align}
\begin{split}
&Q_{i_1}^{1,j=1} = (U^\dag Q(r_1,z_1) U)_{i_1}, \quad Q_{i_1}^{1,j=2} = (U^\dag Q(r_1, z_1) V^*)_{i_1}, \\
&Q_{i_1}^{1,j = 3} = (V^T Q(r_1,z_1) U)_{i_1},  \quad Q_{i_1}^{1,j=4} = (V^T Q(r_1,z_1) V^*)_{i_1}, 
\end{split}
\end{align}
where interaction channel $1$ corresponds to point $(r_1, z_1)$ in the coordinate space. For the $ph$ interaction, two interaction terms (TV and TPV) contribute each with four components, one time-like and three space-like $\mu = -1, 0,1$. Discretizing equations on the 2D mesh with $N_z^{GH}$ nodes in $z$-direction and $N_r^{GL}$ nodes in radial direction generates  $8 \times N_z^{GH} \times N^{GL}_r$ interaction channels. However, we notice that the summation over interaction channels in Eq. (\ref{eq:interaction_matrix}) includes also a second product with $Q_{\tilde{\pi} \tilde{\nu}}^{c, j}$, with $\tilde{\pi}$ states defined in Eq. (\ref{eq:doubled_basis}).
One can show that the following relations hold
\begin{align}
\begin{split}
    Q_{\tilde{\pi} \tilde{\nu}}^{c, j = 1} &= Q_{{\pi} {\nu}}^{c, j = 4}, \\
    Q_{\tilde{\pi} \tilde{\nu}}^{c, j = 2} &= Q_{{\pi} {\nu}}^{c, j = 3}, \\
\end{split}
\end{align}
from which we observe that the second term in Eq. (\ref{eq:interaction_matrix}) induces mixing of the $j$ indices.
Therefore, the total number of $ph$ channels is $N_{ph} = 2 \times 8 \times N_{z}^{GH} \times N_{r}^{GL}$, and the $Q_j$ matrix has the following form
\begin{align}\label{eq:AXDEF_ph_matrix_element}
\begin{split}
Q_{j = 1}^{1 \ldots N_{ph}} &= \begin{pmatrix}
Q_{j = 1}^{TPV} & Q_{j = 1}^{TV}  & Q_{j = 4}^{TPV} & Q_{j = 4}^{TV} \\
\end{pmatrix}, \\
Q_{j = 2}^{1 \ldots N_{ph}} &= \begin{pmatrix}
Q_{j = 2}^{TPV} & Q_{j = 2}^{TV}  & Q_{j = 3}^{TPV} & Q_{j = 3}^{TV} \\
\end{pmatrix}, \\
Q_{j = 3}^{1 \ldots N_{ph}} &= \begin{pmatrix}
Q_{j = 3}^{TPV} & Q_{j = 3}^{TV}  & Q_{j = 2}^{TPV} & Q_{j = 2}^{TV} \\
\end{pmatrix}, \\
Q_{j = 4}^{1 \ldots N_{ph}} &= \begin{pmatrix}
Q_{j = 4}^{TPV} & Q_{j = 4}^{TV}  & Q_{j = 1}^{TPV} & Q_{j = 1}^{TV} \\
\end{pmatrix},
\end{split}
\end{align}
where $Q^{TPV}$ and $Q^{TV}$ correspond to the isovector-pseudovector and isovector-vector separable interaction matrix [see Eqs. (\ref{eq:AXDEF_separable_TV}) and (\ref{eq:AXDEF_separable_TPV})], respectively. The dimension of each matrix is $4 \times N_z^{GH} \times N_r^{GL}$. 

The $pp$ part of the $Q_j$ matrix has a similar structure
\begin{align}
\begin{split}
Q_{j = 1}^{ N_{ph}+1 \ldots N_{ph} +N_{pp}} &= \begin{pmatrix}
W_{j = 1}  & W_{j = 4} \\
\end{pmatrix}, \\
Q_{j = 2}^{ N_{ph}+1 \ldots N_{ph} +N_{pp}} &= \begin{pmatrix}
W_{j = 2}  & W_{j = 3} \\
\end{pmatrix}, \\
Q_{j = 3}^{ N_{ph}+1 \ldots N_{ph} +N_{pp}} &= \begin{pmatrix}
W_{j = 3}  & W_{j = 2} \\
\end{pmatrix}, \\
Q_{j = 4}^{ N_{ph}+1 \ldots N_{ph} +N_{pp}} &= \begin{pmatrix}
W_{j = 4}  & W_{j = 1} \\
\end{pmatrix},
\end{split}
\end{align}
with each submatrix $W_{j}$ of dimension $(3-2T) \times N_r \times N_z$, where $N_z$ and $N_r$ denote the number of harmonic oscillator shells in the $z$- and $r$-directions [cf. Sec. \ref{sec:AXDEF_pp}]. The number of $pp$ interaction channels is given by $N_{pp} = 2 \times (3-2T)\times N_r \times N_z$, and depends on the value of the total isospin $T$. This means that the total number of interaction channels is
\begin{equation}
N_{c} = 2 \times 8 \times N_z^{GH} \times N_r^{GL} + 2 \times (3-2T) \times N_r \times N_z.
\end{equation}
The $N(\omega)$ matrix is diagonal in the q.p. space and has the form 
\begin{align}\label{eq:N_matrix}
\begin{split}
N(\omega)_{j = 1} &= \text{diag}\begin{pmatrix}
\frac{f_{\nu_1} - f_{\pi_1}}{\omega - E_{\pi_1} + E_{\nu_1} + i \eta}, & \ldots,
\end{pmatrix},\\
N(\omega)_{j = 2} &= \text{diag}\begin{pmatrix}
\frac{f_{\tilde{\nu}_1} - f_{\pi_1}}{\omega - E_{\pi_1} + E_{\tilde{\nu}_1} + i \eta}, & \ldots, 
\end{pmatrix}, \\
N(\omega)_{j = 3} &= \text{diag}\begin{pmatrix}
\frac{f_{\nu_1} - f_{\tilde{\pi}_1}}{\omega - E_{\tilde{\pi}_1} + E_{\nu_1} + i \eta}, & \ldots, 
\end{pmatrix},\\
N(\omega)_{j = 4} &= \text{diag}\begin{pmatrix}
\frac{f_{\tilde{\nu}_1} - f_{\tilde{\pi}_1}}{\omega - E_{\tilde{\pi}_1} + E_{\tilde{\nu}_1} + i \eta}, & \ldots, 
\end{pmatrix},
\end{split}
\end{align}
where occupation factors $f_\mu$ and q.p. energies $E_\mu$ are defined in Eq. (\ref{eq:qp_energies_and_occup}). Therefore, at zero-temperature, only $j = 2$ and $j = 3$ components contribute, and have the form
\begin{align}
\begin{split}
N(\omega)_{j = 2} &= \text{diag}\begin{pmatrix}
\frac{1 }{\omega - E_{\pi_1} - E_{{\nu}_1} + i \eta}, & \ldots, 
\end{pmatrix}, \\
N(\omega)_{j = 3} &= \text{diag}\begin{pmatrix}
\frac{ -1}{\omega + E_{{\pi}_1} + E_{\nu_1} + i \eta}, & \ldots, 
\end{pmatrix}.\\
\end{split}
\end{align}
Next, we have to calculate the unperturbed $R^0_{FF}$ response, defined as
\begin{equation}\label{eq:AXDEF_r0ff_matrix}
R^0_{FF}(\omega) = \sum \limits_{j = 1}^4  F_{j}^T   N(\omega)_j  F_{j},
\end{equation}
 where $R_{FF}^0(\omega) \in \mathbb{C}$, and $F_j \in \mathbb{R}^{N_{pair}}$. Finally, the $R_{cF}^0$ reduced response reads
 \begin{equation}\label{eq:AXDEF_r0cf_matrix}
R^0(\omega)_{cF} = \sum_{j=1,4} Q^T_{c,j}   N(\omega)_j  F_j,
\end{equation}
of the dimension $N_{c}$, mixing both the residual interaction and the external field matrix element. The $v_{c c^\prime}$ interaction matrix in Eq. (\ref{eq:interaction_matrix}) is diagonal (since we consider no derivative terms in the residual interaction) and is given by a direct sum of diagonal matrices containing interaction couplings

\begin{equation}
v_{cc^\prime} = \bigoplus_{i = 1}^{16+2(3-2T)} v^{i}_{c c^\prime},
\end{equation}
where for the $ph$ channels
\begin{align}
    \begin{split}
        &v^{(1)} = v^{(9+(3-2T))} = +g_0/(2\pi)\mathbb{I}_{N_z^{GH} \times N_r^{GL}},  \\
        &v^{(2)} = v^{(10+(3-2T))} = -g_0/(2\pi)\mathbb{I}_{N_z^{GH} \times N_r^{GL}}, \\
        &v^{(3)} = v^{(11+(3-2T))} = -g_0/(2\pi)\mathbb{I}_{N_z^{GH} \times N_r^{GL}}, \\ 
        &v^{(4)} = v^{(12+(3-2T))} = -g_0/(2\pi)\mathbb{I}_{N_z^{GH} \times N_r^{GL}}, \\
        &v^{(5)} = v^{(13+(3-2T))} =  +\alpha_{TV}/(2\pi)\mathbb{I}_{N_z^{GH} \times N_r^{GL}},\\
        &v^{(6)} = v^{(14+(3-2T))} =  -\alpha_{TV}/(2\pi)\mathbb{I}_{N_z^{GH} \times N_r^{GL}},\\
        &v^{(7)} = v^{(15+(3-2T))} =  -\alpha_{TV}/(2\pi)\mathbb{I}_{N_z^{GH} \times N_r^{GL}},\\
        &v^{(8)} = v^{(16+(3-2T))} =  -\alpha_{TV}/(2\pi)\mathbb{I}_{N_z^{GH} \times N_r^{GL}},\\
    \end{split}
\end{align}
with the total number of channels depending on the isospin $T$. For the pairing channels we take the average of the separable pairing strength for neutrons ($G_n$) and protons ($G_p$), so that their form is
\begin{align}
\begin{split}
    &v^{(9)} = v^{(18)} = (G_n+G_p)/2 \mathbb{I}_{N_z \times N_r}, \quad T = 1, \\
    &v^{(9,10,11)} = v^{(20,21,22)} = V_0^{pp}(G_n+G_p)/2 \mathbb{I}_{N_z \times N_r}, \quad T = 0.\\
\end{split}
\end{align}
After all the necessary matrices are calculated, we first invert the matrix $\delta_{c c^\prime} - \sum \limits_{c^{\prime \prime}} R^0_{c c^{\prime \prime}} v_{c^{\prime \prime} c^\prime}$ and then calculate the $R_{cF}$ response
\begin{equation}\label{eq:AXDEF_rcf_calculation}
R_{cF}(\omega) = \sum \limits_{c^\prime}[ \delta_{c c^\prime} - \sum \limits_{c^{\prime \prime}} R^0_{c c^{\prime \prime}}(\omega) v_{c^{\prime \prime^\prime} c}]^{-1} R_{c^\prime F}^0(\omega).
\end{equation}
Finally, the response function is obtained as
\begin{equation}
R_{FF}(\omega) = R_{FF}^0(\omega) +  \sum \limits _{c c^\prime} R^0_{cF}(\omega)v_{c c^\prime} R_{c^\prime F}(\omega),
\end{equation}
its imaginary part giving the strength function [see Eq. (\ref{eq:strength_function_Im})].

\begin{figure}
    \centering
    \includegraphics[width = \linewidth]{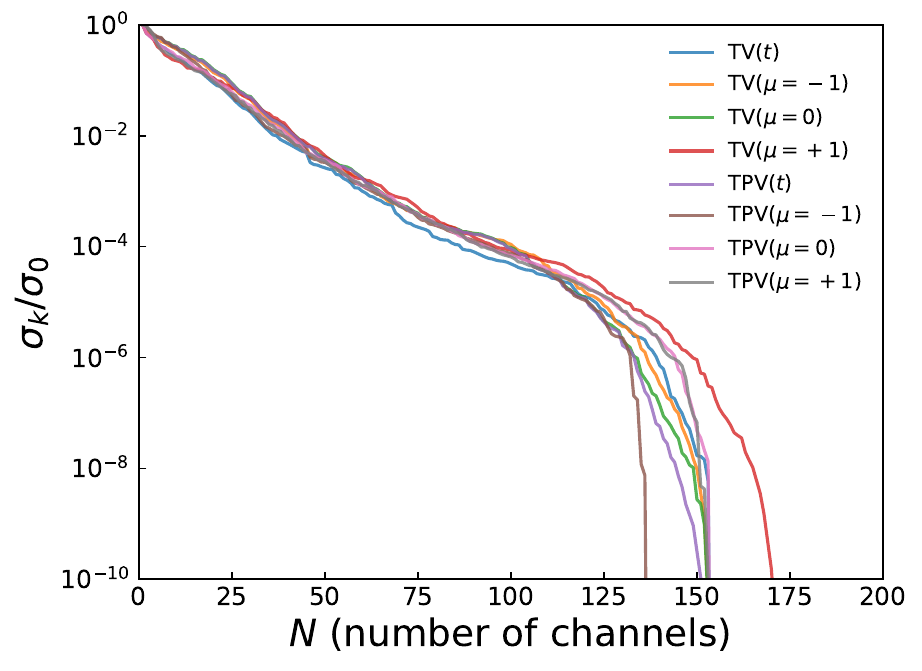}
    \caption{The decay of singular values $\sigma_k/\sigma_0$ for different interaction channels of the $Q_j$ matrix. The results are shown for $j = 2$ component.}
    \label{fig:Q_cut}
\end{figure}

Therefore, from the computational perspective, for a given energy $\omega$, one has to multiply matrices of the size $N_{c} \times N_{pair}$ in Eq. (\ref{eq:AXDEF_r0_matrix}), vectors of dimension $N_{pair}$ in Eq. (\ref{eq:AXDEF_r0ff_matrix}) and their cross-term in Eq. (\ref{eq:AXDEF_r0cf_matrix}), all for $j = 1,2,3$, and 4 (or $ j = 2,3$ at zero-temperature).  After that, the square matrix $R^0_{c c^\prime}$ of the size $N_c \times N_c$ has to be inverted and multiplied with $R^0_{cF}$ in Eq. (\ref{eq:AXDEF_rcf_calculation}), other operations being less computationally expensive. To illustrate, if we use $N_{osc} = 16$ h.o. shells, then $N_z^{GH} = N_r^{GL} \sim 16$ for the mesh and $N_z = N_r \sim 16$ for the pairing interaction. Therefore, $N_{ph} \sim 4096$ and $N_{pp} \sim 1500$ for the more expensive isoscalar pairing (500 for isovector). The total number of channels is $N_{c} \sim 5600$. The number of pairs for the $K = 0$ mode of the GT transitions is $N_{pair} \sim 50000$, meaning that the largest matrix size for the multiplication is of the order $5600 \times 50000$ and for matrix inversion $5600 \times 5600$, easily manageable on a moderate computer cluster. These dimensions can be further reduced as explained in Appendix \ref{sec:appb}.

\begin{figure}[t!]
    \centering
    \includegraphics[width=\linewidth]{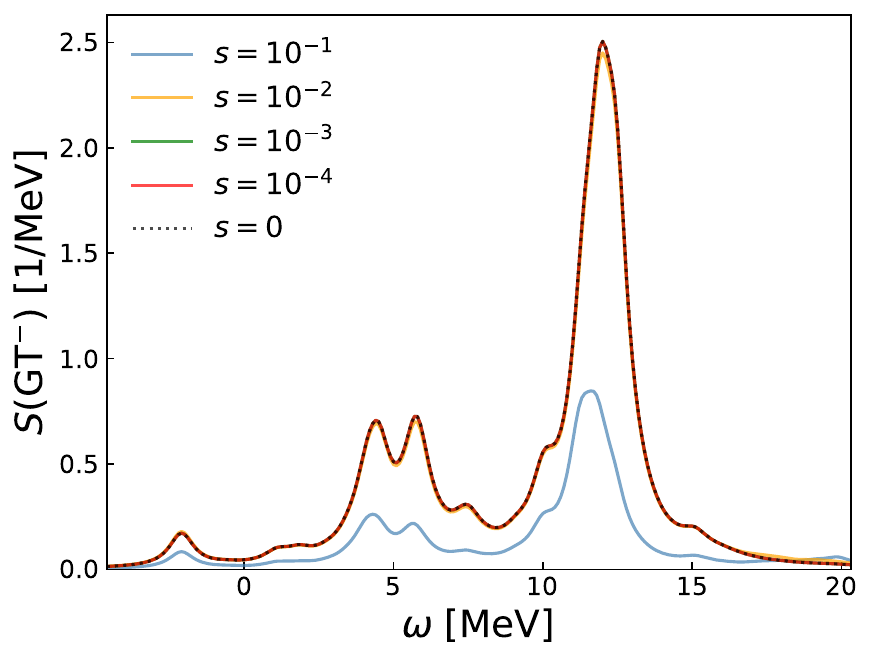}
    \caption{The $K=1$ mode of GT${}^-$ strength in ${}^{62}$Fe calculated with a different truncation on the power spectrum $s$.}
    \label{fig:SVD}
\end{figure}

\section{Speeding-up calculation of unperturbed response matrices with Singular Value Decomposition}\label{sec:appb}
A significant part of the computational time is spent on calculating the unperturbed reduced response function 
\begin{equation}\label{eq:reduced_unp_response}
    R^0_{c c^\prime}(\omega) = \sum \limits_{j=1}^4 \sum \limits_{\pi \nu } (Q^{c, j}_{\pi \nu})^* N(\omega)_{\pi \nu,j} Q^{c^\prime, j}_{\pi \nu},
\end{equation}
where the $N(\omega)$ matrix is defined in Eq. (\ref{eq:N_matrix}). Since the dimension of the $Q_{\pi \nu}^{c,j}$ matrix ($N_{pair} \times N_c$) becomes quite large for reasonable basis size, it is worthwhile to investigate the rank of this matrix. We perform the singular value decomposition (SVD)
\begin{equation}
    Q_j = U_j \Sigma_j  V^T_j, \quad j = 1,\ldots,4,
\end{equation}
where the diagonal $\Sigma_j \in \mathbb{R}^{N_{pair} \times N_C}$ matrix contains the singular values, while $U_j \in \mathbb{C}^{N_{pair} \times N_{pair}}$, and $V^T_j \in \mathbb{C}^{N_c \times N_c}$ are matrices with left and right singular vectors. Therefore, in the matrix notation (suppressing q.p. pair and channel indices) we can rewrite Eq. (\ref{eq:reduced_unp_response}) as
\begin{align}\label{eq:R0_matrix}
    \begin{split}
        R^0 = \sum \limits_{j =1}^4 V_j \Sigma_j [U^T_j N(\omega)_j U_j] \Sigma_j V^T_j.
    \end{split}
\end{align}

If the singular value spectrum decays fast, so that all singular values $\sigma_k$ become smaller than some predefined threshold (e.g. $\sigma_k/\sigma_1 \leq 10^{-8}$) we can use the so-called low-rank approximation to write $Q_j$ matrix by using only the first $N_k \leq N_{c}$ columns of $U_j$ matrix, and $N_k$ rows of $V^T_j$ matrix
\begin{equation}
    Q_j \approx \hat{U}_{N_{pair} \times N_k} \times \hat{\Sigma}_{N_k \times N_k} \times \hat{V}^T_{N_k \times N_c},
\end{equation}
where subscripts denote the matrix dimension. In the low-rank approximation Eq. (\ref{eq:R0_matrix}) has the form
\begin{equation}
    R^0 \approx \sum \limits_{j = 1}^4 \hat{V}_j \hat{\Sigma}_j[\hat{U}^T_j N(\omega)_j \hat{U}_j] \hat{\Sigma}_j \hat{V}^T_j,
\end{equation}
where the inner multiplication now requires much less computational effort. To test our approach, we study the singular value decay of the $Q_j$ matrix in the ${}^{62}$Fe isotope for the $K=1$ GT${}^-$ transition. We use $N_{osc} = 16$ harmonic oscillator shells, and quadrupole deformation constrained to $\beta_2 = +0.2$. Results are shown in Fig. \ref{fig:Q_cut} for $j = 2$ component (analogous results follow for other $j$ components). Although the full dimension is $N_c = 361$, we observe that this can be reduced by more than 50\% for all interaction channels.

\begin{table}[t!]
    \centering
    \caption{The ratio between the cut-off number of channels $N_k$ [determined by the power spectrum $s$ defined in Eq. (\ref{eq:power-spectrum})] and total number of channels $N_c$ (in \%) for different terms of isovector-vector (TV) and isovector-pseudovector (TPV) interaction. The total number of channels is $N_c = 361$, and $j = 2$.}
    \begin{tabular}{c|cccc}
    \hline
    \hline
       &  TV(t) & TV($\mu = -1$) &  TV($\mu = 0$) & TV($\mu = +1$) \\
       \hline
     $s = 10^{-1}$ & 5.5 & 6.1 & 6.1 & 6.6 \\
     $s = 10^{-2}$ & 11.1 & 11.4 & 11.6 & 12.7 \\
     $s = 10^{-3}$ & 18.0 & 19.1 & 18.6 & 20.5 \\
     $s = 10^{-4}$ & 27.4 & 28.3 & 27.4 & 30.5 \\
     \hline
       &  TPV(t) & TV($\mu = -1$) &  TV($\mu = 0$) & TV($\mu = +1$) \\
       \hline
        $s = 10^{-1}$ & 6.1 & 5.3 & 6.1 & 5.8 \\
     $s = 10^{-2}$ & 11.4 & 11.1 & 11.6 & 11.6 \\
     $s = 10^{-3}$ & 18.6 & 18.6 & 19.4 & 19.4 \\
     $s = 10^{-4}$ & 27.1 & 27.1 & 28.8 & 28.3 \\
       \hline
    \end{tabular}
    \label{tab:cutoff}
\end{table}

In order to define where to truncate $N_k$, we inspect the power spectrum defined as
\begin{equation}
\label{eq:power-spectrum}
    s = 1 - \frac{\sum \limits_{i = 1}^{N_k} \sigma_i}{\sum \limits_{i = 1}^{N_c} \sigma_i}.
\end{equation}
In Fig. \ref{fig:SVD} we show the strength function for our test case using different values of $s$ ranging from $10^{-1}$ ($99\%$ of the spectrum is retained) to $10^{-4}$ (99.999\% of the spectrum is retained). We observe that already for $s = 10^{-2}$ the strength is well converged, nearly indistinguishable from the full strength without any approximation ($s = 0$). In Tab. \ref{tab:cutoff}, we show the percentage of the retained number of interaction channels for various values of the power spectrum $s$. We note that using $s = 10^{-2}$ corresponds to retaining around 11\% of the total number of interaction channels $N_c$. Since the computational effort of matrix-matrix multiplication scales with matrix dimension as $N^3$, such a cut-off leads to significant speed-up of the code.

\bibliographystyle{apsrev4-2}
\bibliography{bibl}

\end{document}